\documentstyle[aps,epsfig]{revtex}
\renewcommand{\thefootnote}{\fnsymbol{footnote}}
\begin{document}
\setcounter{footnote}{0}
\begin{center}
{\Large\bf Gluon self-energy in a two-flavor color superconductor}
\\[1cm]
Dirk H.\ Rischke
\\ ~~ \\
{\it Institut f\"ur Theoretische Physik}\\
{\it Johann Wolfgang Goethe -- Universit\"at} \\
{\it Robert-Mayer-Str.\ 8--10, D-60054 Frankfurt/Main, Germany}\\
{\it email: drischke@th.physik.uni-frankfurt.de}
\\ ~~ \\ ~~ \\
\end{center}
\begin{abstract} 
The energy and momentum dependence of the gluon self-energy
is investigated in a color superconductor with two flavors of massless quarks.
The presence of a color-superconducting quark-quark condensate 
modifies the gluon self-energy for energies 
which are of the order of the gap parameter.
For gluon energies much larger than the gap,
the self-energy assumes the form given
by the standard hard-dense loop approximation.
It is shown that this mo\-dification of the gluon self-energy
does not affect the magnitude of the gap
to leading and subleading order in the weak-coupling limit.
\end{abstract}
\renewcommand{\thefootnote}{\arabic{footnote}}
\setcounter{footnote}{0}

\section{Introduction}

Single-gluon exchange between two quarks
is attractive in the color-antitriplet channel.
Therefore, sufficiently cold and dense quark matter is a color 
superconductor \cite{bailinlove}. 
When the quark-chemical
potential $\mu \gg \Lambda_{\rm QCD}$, asymptotic freedom 
\cite{asympfreed} implies
that the strong coupling constant $g$ at the scale
$\mu$ is much smaller than unity, $g(\mu) \ll 1$. This allows
a controlled calculation of the color-superconducting gap parameter
$\phi$ in the weak-coupling limit.

In order to compute the gap parameter, one has to solve a gap equation.
In general, the gap equation determines the gap parameter $\phi(K)$ 
as a function of 4-momentum $K^\mu \equiv (k_0,{\bf k})$
of the fermionic quasiparticles. However, of physical interest
is only the value of the gap function {\em on the quasiparticle
mass shell} \cite{rdpdhr2}, {\it i.e.}, for
$k_0 \equiv \epsilon_k$, where
$\epsilon_k \equiv \sqrt{(\omega^0_k- \mu)^2 + \phi^2}$ is
the fermionic quasiparticle excitation energy. Here, $\omega^0_k$ is the
kinetic energy of non-interacting particles, $\omega^0_k = k
\equiv |{\bf k}|$ in the ultrarelativistic case.
Then, at zero temperature, $T=0$,
the gap equation is typically of the form
\begin{equation} \label{genericgapequation}
\phi(\epsilon_k,{\bf k}) = g^2 \, \rho(\mu)
\int \frac{{\rm d}q}{\epsilon_q} 
\int {\rm d} \cos \theta \;
\Delta (\epsilon_k-\epsilon_q,{\bf k} - {\bf q}) \; 
\phi(\epsilon_q,{\bf q}) \,\, .
\end{equation}
Here, $\rho(\mu)$ is the density of states at the
Fermi surface, $\rho(\mu) \sim \mu^2$ in the ultrarelativistic
case. Furthermore,
it is assumed that the attractive interaction between fermions
is mediated by boson exchange, with $g$ being the fermion-boson
coupling constant and $\Delta(P)$ the boson propagator;
$\theta \equiv \arccos ({\bf k}, {\bf q})$ is the angle 
between the 3-momenta of in- and out-going fermions
in a scattering process. 

For an ordinary superconductor, standard BCS theory \cite{BCS,fetter} generally
assumes that the attractive phonon-exchange interaction between electrons is
local, {\it i.e.}, a point-like four-fermion interaction, and isotropic, 
$\Delta (P) \equiv 1/\Lambda^2 $, where $\Lambda$ is a
constant with the dimension of energy.
As a consequence, the integral over $\cos \theta$ 
in Eq.\ (\ref{genericgapequation}) is trivial.
In a superconductor, the maximum phonon momentum is $p_D$, the
Debye momentum.
In weak coupling, $p_D \sim g \mu \ll \mu$.
Consequently, the momentum exchange between
electrons in a phonon-mediated scattering process is limited to the range 
$|{\bf k}-{\bf q}| \leq  p_D$.
Let us consider $k=k_F$, where $k_F$ is the Fermi momentum, 
and denote the value of the gap function at the
Fermi surface by $\phi$.
The restriction on possible phonon momenta then translates
to limiting the $q$-integration
to the (in weak coupling, narrow) region
$k_F - p_D \leq q \leq k_F + p_D$ around the Fermi surface. Assuming
that the momentum dependence of $\phi(\epsilon_q,{\bf q})$
near the Fermi surface, $q \simeq k_F$, 
is weak, $\phi (\epsilon_q,{\bf q}) \simeq \phi = const.$, the 
gap equation simplifies to
\begin{equation} \label{BCSgapequation}
\phi = \frac{g^2}{c_{\rm BCS}} \; 
\ln \left( \frac{\omega_D + \sqrt{\omega_D^2 + \phi^2}}{\phi} \right) 
\; \phi\,\, ,
\end{equation}
where $1/c_{\rm BCS} \sim \rho(\mu)/\Lambda^2$. Here,
$\omega_D \equiv v_F \, p_D$ is the Debye frequency and
$v_F$ the Fermi velocity, $v_F = 1 $ in the ultrarelativistic limit.
For any, even arbitrarily weak, attractive interaction there
is always a nontrivial solution $\phi \neq 0$ to this equation.
Consequently, the constant
factor $\phi$ common to both sides can be divided out.
In weak coupling, $g^2 \ll 1$, and in order to solve Eq.\ 
(\ref{BCSgapequation})
the logarithm has to be $\sim 1/g^2 \gg 1$, 
to compensate the small prefactor $g^2 \ll 1$. This
requires $\phi \ll \omega_D$. Then one may approximate
$\ln [(\omega_D + \sqrt{\omega_D^2 + \phi^2})/\phi] \simeq
\ln(2\, \omega_D/\phi)$, and the solution is
\begin{equation} \label{BCSresult}
\phi = b_{\rm BCS} \; \mu \;\exp \left(- \frac{c_{\rm BCS}}{g^2}
\right) \,\, ,
\end{equation}
where $b_{\rm BCS} \equiv 2\,\omega_D/\mu$ is a dimensionless constant.
This discussion shows that the constant $c_{\rm BCS}$ in the exponent is
determined by the constant {\em in front\/} of the ``BCS logarithm''
$\int {\rm d}q/\epsilon_q \sim \ln(b_{\rm BCS}\, \mu/\phi)$, while
the constant in the prefactor, $b_{\rm BCS}$, is determined by
the constant ``under'' the BCS logarithm, in this case
the size of the $q$-integration region.

The discussion can be easily generalized to non-local interactions
of finite range, for instance massive scalar boson exchange
\cite{rdpdhrscalar}. In this case,
$\Delta(P) \equiv (M_B^2 - P^2)^{-1}$ is the boson propagator,
with $M_B$ the boson mass, $P^2 \equiv p_0^2 - {\bf p}^2$. 
For the following, let us assume that fermions are massless, and
that the boson mass is generated by in-medium effects, $M_B \simeq g \mu$,
{\it i.e.}, in weak coupling the different energy scales in the problem
are ordered, $\phi \ll M_B \ll \mu$.
The angular integral is no longer
trivial; in the ultrarelativistic case and near the Fermi surface,
$k \simeq q \simeq k_F \equiv \mu$,
\begin{equation} \label{collinearfactor}
\rho(\mu) \, \Delta(P) \sim \frac{1}{2\, (1 - \cos \theta)+M_B^2/\mu^2} \,\, ,
\end{equation}
where the energy dependence of the boson propagator
has been neglected, since $p_0 \equiv \epsilon_k - \epsilon_q \sim \phi 
\ll M_B$.
Obviously, the term (\ref{collinearfactor}) enhances the contribution
from small-angle fermion-fermion scattering, $\cos \theta \simeq 1$, 
to the gap equation (\ref{genericgapequation}).
Performing the angular integral introduces an additional logarithm,
in the following called ``collinear logarithm'',
in comparison to Eq.\ (\ref{BCSgapequation}),
\begin{equation} \label{massivebosonexchangegapequation}
\phi = \frac{g^2}{c_{\rm B}} \; 
\ln \left( \frac{b_{\rm B}\, \mu}{\phi} \right) \; 
\ln \left(\frac{2 \, \mu}{M_B} \right)
\; \phi\,\, ,
\end{equation}
where I exploited the hierarchy of scales in weak coupling.
The solution is rather similar to the one in BCS theory, Eq.\ 
(\ref{BCSresult}),
\begin{equation} \label{MBXresult}
\phi = b_{\rm B}\, \mu \, \exp \left( - \frac{c_{\rm B}}{g^2 \, \ln 
(2\,\mu/M_B)}
\right)\,\, .
\end{equation}
Quantitatively, the difference to Eq.\ (\ref{BCSresult})
is that, in weak coupling, the gap becomes larger, since the
coupling constant $g^2$ is replaced by $g^2 \ln (2\,\mu/M_B)
\sim g^2 \ln(1/g) \gg g^2$.

In QCD, gluon exchange is also a non-local interaction.
The difference to the previous case is that gluons are massless
and thus the interaction has infinite range.
In the vacuum, this holds for electric as well as magnetic gluons.
In a dense medium, however, electric and non-static magnetic
gluon exchange is screened. Essentially, these gluons behave
like the scalar bosons of the previous example with $M_B \equiv m_g$,
where
\begin{equation} \label{gluonmass}
m_g = \sqrt{\frac{N_f}{6 \pi^2}} \; g \, \mu
\end{equation}
is the gluon mass at $T=0$, with $N_f$ being 
the number of massless quark flavors participating
in screening electric and non-static magnetic gluons.
Screening reduces the range of the interaction
to distances $< m_g^{-1}$.
On the other hand, the exchange of almost static
magnetic gluons is still unscreened and thus of infinite range.

In the vacuum, the gluon
propagator is $\Delta(P) \sim - 1/P^2$, and the previously
discussed collinear
enhancement in the gap equation is much stronger: 
instead of being cut off by a mass term $M_B$,
it is cut off by the gluon {\em energy}, 
$p_0 = \epsilon_k - \epsilon_q \sim \phi \ll M_B \simeq g \mu$.
(In the previous case, the boson energy was negligible compared
to its mass.)
The collinear logarithm in the gap equation is now
$\sim \ln [2\,\mu/|\epsilon_k - \epsilon_q|]
\sim \ln (2\,\mu/\phi)$. It is thus of the
same order as the BCS logarithm,
and consequently much larger than the collinear
logarithm $\ln(2\,\mu/M_B) \sim \ln(1/g)$ in the previous case.
In a dense medium, the gluon propagator is
more complicated, but since
the exchange of almost static magnetic gluons 
is unscreened, it still
gives rise to the same collinear logarithm $\ln (2\,\mu/\phi)$
as in the vacuum \cite{rdpdhr2}.

In QCD, there is no restriction on the magnitude of the gluon momentum
in the gap equation. Nevertheless, in weak coupling the gap
function $\phi(\epsilon_q,{\bf q})$ is strongly peaked at the
Fermi surface $q = k_F$ \cite{rdpdhr2,son,schaferwilczek}.
Effectively, this again restricts the $q$-integration
to a narrow range $k_F-\delta \leq q \leq k_F + \delta$
around the Fermi surface, $\phi \ll \delta \ll \mu$. For an order-of-magnitude
estimate (but not for a numerically correct solution), in this range
one may take $\phi(\epsilon_k, {\bf k}) \simeq 
\phi (\epsilon_q, {\bf q}) \simeq \phi = const.$, 
such that the gap equation assumes the form
\begin{equation} \label{QCDgapequation}
\phi = \frac{g^2}{c^2} \, \left[\;
\ln^2 \left( \frac{2 \, \delta}{\phi} \right) 
+  b'\, \ln \left( \frac{2\, \delta}{\phi} \right) +  d\;
\right] \; \phi\,\, .
\end{equation}
The first term contains two powers of the logarithm 
$\ln (2\, \delta/\phi)$. The first one is
the BCS logarithm. The second one is
the collinear logarithm arising from the exchange of
massless gluons. (Any constant under this logarithm different
from $2\,\delta$ has been absorbed in $b'$.)
The value of the constant $c^2$ is determined by how many gluon
degrees of freedom are unscreened and thus give rise
to a collinear logarithm in addition to the BCS logarithm.
In vacuum, these are all gluons, while in medium, only 
almost static magnetic gluons contribute to $c^2$.
The second term contains only one power of the logarithm 
$\ln(2\, \delta/\phi)$: the BCS logarithm.
In a strongly interacting, dense medium, 
this logarithm arises from exchange of screened electric and 
non-static magnetic gluons \cite{rdpdhr2}. The value of $b'$
is determined by the number of degrees of freedom of such gluons. 
As seen above,
screened boson exchange also gives rise to a collinear logarithm, but
this time of the form $\ln (2\, \mu/M_B)$, $M_B \equiv m_g$ for gluons.
This logarithm has been absorbed in the constant $b'$.
Finally, all contributions without any logarithm have been
summed up into the constant $d$.

In weak coupling, $\phi \ll \delta$ and $\ln (2\, \delta/\phi) \gg 1$. Thus,
all unscreened gluons contribute to
{\em leading order\/} to the gap equation, {\it i.e.}, to the first term
in Eq.\ (\ref{QCDgapequation}), while all screened
gluons contribute to {\em subleading order}, {\it i.e.}, to the
second term. All other contributions make up the third,
{\em sub-subleading\/} order term.

The solution of Eq.\ (\ref{QCDgapequation}) is straightforward,
\begin{equation} \label{barroisresult}
\phi = b\, \mu \, \exp\left(- \frac{c}{g}\right) \;
\left[ 1 + O(g) \right]\,\, ,
\end{equation}
where $b \,\mu \equiv 2\, \delta\, \exp (b'/2)$.
A more thorough analysis \cite{rdpdhr2} shows that
$b$ is actually independent of $\delta$, and solely determined
by $b'$, {\it i.e.}, by the contributions of
screened gluons and the constants under the collinear logarithm. 
The constant $d$ does not appear to leading
and subleading order in the solution of the gap equation;
it is part of the terms of order $O(g)$
on the right-hand side of (\ref{barroisresult}).

Observe that the power of the coupling $g$ in the exponent
is reduced as compared to the BCS result (\ref{BCSresult})
or the result (\ref{MBXresult}) for massive boson exchange.
This was first noted by Barrois \cite{barrois}, but
never made it into the published literature.
Barrois assumed that the gluon propagator is the same as
in vacuum and consequently, as discussed above, all
gluon degrees of freedom are unscreened.

In a dense medium, however, only almost static magnetic gluons are
unscreened, while electric and non-static magnetic gluons are
screened. 
In a seminal paper, Son \cite{son} derived the result
(\ref{barroisresult}) with the gluon propagator in
the so-called ``hard-dense-loop'' (HDL) limit
\cite{LeBellac,finitedens}, and obtained
\begin{equation} \label{sonresult1}
c = \frac{3 \pi^2}{\sqrt{2}}\,\, .
\end{equation}
Son also gave an estimate for the constant $b$,
\begin{equation} \label{sonresult2}
b = \frac{b_0}{g^5} \,\, ,
\end{equation}
with some constant $b_0$ of order one.
The parametric dependence of $b$ on $g$ arises from 
the gluon mass, Eq.\ (\ref{gluonmass}). Several authors 
\cite{rdpdhr2,schaferwilczek,hongetal,hsuschwetz}
have confirmed the results (\ref{sonresult1}) and (\ref{sonresult2}) 
and refined the estimate for $b_0$,
\begin{equation} \label{b}
b_0 = 512 \; \pi^4 \left(\frac{2}{N_f}\right)^{5/2} b_0'\,\, ,
\end{equation}
where the dependence on $N_f$ arises from that of
the gluon mass (\ref{gluonmass}), and
$b_0'=1$ under the present approximations.

In a dense medium, not only the gluon propagator is modified.
Brown, Liu, and Ren \cite{rockef} 
included a finite, $\mu$-dependent 
contribution to the quark wavefunction renormalization
in their estimate for $b_0'$,
\begin{equation} \label{rockefresult}
b_0' = \exp\left(-\frac{\pi^2+4}{8}\right) \, b_0'' \simeq 0.176 \; b_0''\,\, .
\end{equation}
They also asserted that there
are no further corrections to $b_0'$ at this order in $g$,
{\it i.e.}, $b_0'' = 1 + O(g)$.

Beane, Bedaque, and Savage \cite{beane} argued that the
coupling constant in the gap equation should not be taken
at the scale $\mu$, but at the scale of the momentum of the
exchanged gluon. Using the standard running of the coupling
constant in the vacuum they obtained $b_0'' = \exp[33(\pi^2-4)/64]
\simeq 20$.
At present, it remains an interesting open problem to 
check the validity of their arguments by systematically computing vertex
corrections to the gap equation. Furthermore, their estimate could
be improved replacing the vacuum running of the coupling by that in a dense
medium.

Sub-subleading contributions, {\it i.e.}, of order $g$ in the constant $b_0''$,
or equivalently, contributions to the constant $d$ in Eq.\ 
(\ref{QCDgapequation}),
can arise from a variety of effects, for instance from the
finite lifetime of quark excitations away from the Fermi surface \cite{manuel}.
Also, an apparent gauge dependence of the result (\ref{barroisresult}) 
surfaces at this order \cite{shusterrajagopal}. However, the gap on the 
quasiparticle mass shell is a physical observable and thus in
principle gauge-independent. This indicates that, for a
complete solution to sub-subleading order, one has to go
beyond the one-boson exchange approximation in the gap equation.
This task appears, at least at present, too difficult to be feasible.

In this paper, I focus instead on a contribution which is
potentially of leading or subleading order. 
At small temperatures $T \sim \phi \ll \mu$, 
the dominant contribution
to the one-loop gluon self-energy
comes from a quark loop; it is $\sim g^2 \mu^2$, while 
gluon (and ghost) loops contribute a term 
$\sim g^2 T^2$ and are thus suppressed \cite{dhr2f}. 
In the standard HDL approximation, however, the quark excitations 
in the quark loop are taken to be those in
a {\em normal\/} and not a superconducting medium.
This is in principle inconsistent, as the fermion excitation
spectrum in a superconductor differs from that in a
normal conductor \cite{rdpdhrscalar}.
In Refs.\ \cite{dhr2f,dhr3f} I argued that using the
correct gluon self-energy
in the gap equation could in principle affect the value
of the color-superconducting gap \cite{hormuzhsu}. 
Consequently, I derived general
expressions for the gluon self-energy in the two-
and three-flavor case, and computed the self-energy in the
limit of vanishing energy and momentum, which yields the Debye and Meissner
masses in a color-superconductor.

It turned out that in the two-flavor case, where
the color-superconducting condensate breaks the $SU(3)_c$ color
gauge symmetry to $SU(2)_c$, the three gluons corresponding to the unbroken
$SU(2)_c$ subgroup remain massless, like in the vacuum \cite{dhr2f}.
Naively, if the propagator of these gluons is the same
as in vacuum also at {\em nonzero\/} energy $p_0$ and momentum ${\bf p}$, 
this would change the number of modes mediating long-range interactions.
Remembering the above discussion of the gap equation
(\ref{QCDgapequation}), this would affect
the leading-order contribution to the gap equation and
could in principle change the value (\ref{sonresult1}) for the
coefficient $c$. 

It turns out, however, that
for zero gluon energy $p_0 = 0$ and gluon momenta $|{\bf p}| \equiv p 
\gg \phi$,
the self-energies approach their HDL form \cite{dhr2f}. Moreover,
for energies and momenta much smaller than the gap, $p_0,\, p \ll \phi$,
these gluons do not behave exactly like in vacuum, but like in a 
medium with a large dielectric constant \cite{rss}. 
A change of the leading-order contribution to the gap equation, 
and thus of the coefficient $c$, appears therefore unlikely, but a change
of the subleading contribution cannot be excluded.
In order to decide this question, however,
knowledge of the gluon self-energy in the limited
range of energies and momenta $p_0,\, p \ll \phi$, or
for $p_0 = 0$ and $p \gg \phi$ is insufficient;
one has to compute
the gluon self-energy for energies and momenta of 
relevance for the gap equation (\ref{genericgapequation}).
As the integral on the right-hand side of (\ref{genericgapequation})
is dominated by quark energies and momenta close to the Fermi
surface, the range of relevant gluon energies and momenta
is $p_0 ,\, p \ll \mu$, but not the more restrictive range
$p_0,\, p \ll \phi$ or the special case $p_0 = 0,\, p \gg \phi$.
In the present paper, I therefore compute the gluon self-energy 
for $p_0 ,\, p \ll \mu$ in a two-flavor color superconductor.

The gap equation in the two-flavor case was derived in \cite{rdpdhr2}.
Assuming that Cooper pairs are formed from quarks with fundamental
colors 1 and 2, such that the condensate has fundamental color
(anti-) 3, one obtains
\begin{equation} \label{gapequation}
\Phi^+(K) = \frac{3}{4} \, 
g^2 \frac{T}{V} \sum_Q \gamma_\mu \, \left[ \Delta^{\mu \nu}_{11}(K-Q)
- \frac{1}{9}\, \Delta^{\mu \nu}_{88} (K-Q) \right] \, 
\Xi^+(Q) \gamma_\nu\,\, .
\end{equation}
Here, $\Phi^+$ is the gap matrix in Dirac space,
$V$ is the volume of the system, $\gamma_\mu$ are Dirac matrices,
$\Delta^{\mu \nu}_{ab}$ is the gluon propagator,
and $\Xi^+$ is the off-diagonal element of
the quasiparticle propagator in Nambu--Gor'kov space.

Only gluons mediating
between gapped quarks enter the gap equation, {\it i.e.}, gluons
with adjoint colors 1, 2, 3, and 8. Gluons mediating between
gapped and ungapped quarks, {\it i.e.},
gluons with adjoint colors 4, 5, 6, and 7, do not appear.
Moreover, the propagators for gluons of the 
unbroken $SU(2)_c$ subgroup
are identical, $\Delta^{\mu \nu}_{11} = \Delta^{\mu \nu}_{22} = 
\Delta^{\mu \nu}_{33}$, but different from the propagator for
the eighth gluon, $\Delta^{\mu \nu}_{88}$. 

I work in pure Coulomb gauge, where the gluon propagator for
colors 1, 2, 3, and 8 assumes the
form \cite{dhr2f,GPY}
\begin{mathletters}
\begin{eqnarray}
\Delta^{00}_{ab} (P) & = & - \delta_{ab} \;
 \frac{1}{{\bf p}^2 - \Pi^{00}_{aa}(P)} \,\, , \\
\Delta^{0i}_{ab} (P) & = & 0 \,\, , \\
\Delta^{ij}_{ab} (P) & = & - \delta_{ab} \;
(\delta^{ij} -  \hat{p}^i \, \hat{p}^j) \;
\frac{1}{P^2 - \Pi^t_{aa}(P)}\,\, .
\end{eqnarray}
\end{mathletters}
Here, $\Pi^{00}_{aa}(P)$ is the $00$-component and
\begin{equation}
\Pi^t_{aa}(P) \equiv \frac{1}{2}\; (\delta_{ij} - \hat{p}_i \, \hat{p}_j) \;
\Pi_{aa}^{ij}(P)
\end{equation}
the spatially transverse component of the
gluon self-energy $\Pi^{\mu \nu}_{aa}(P)$, $\hat{\bf p} \equiv {\bf p}/p$.
In the following, I will (somewhat imprecisely) term $\Pi^{00}_{aa}$
the {\em electric\/} self-energy, and
$\Pi^t_{aa}$ the {\em magnetic\/} self-energy.

Apparently, in order to settle the question whether the modification of
the gluon self-energy in a color superconductor (as compared to
the HDL approximation) changes the solution of the gap equation
(\ref{gapequation}), one only has to consider the following four components
of the gluon self-energy: $\Pi^{00}_{11}, \, \Pi^{00}_{88}, \,
\Pi^t_{11}$, and $\Pi^t_{88}$.
The imaginary and real parts of these four components are computed 
in Sec.\ \ref{ii}. In Sec.\ \ref{iii}, I determine the 
spectral densities of electric
and magnetic gluons as a function of 
gluon energy and momentum.
Section \ref{iv} discusses the effect of the modification
of the gluon self-energy in a color superconductor on the
solution of the gap equation. I conclude in Sec.\ \ref{v} with
a summary of the results. Unless mentioned otherwise,
throughout this paper I work at zero temperature, $T=0$.
I use natural units and the metric tensor 
$g^{\mu \nu} = {\rm diag}\, (+,-,-,-)$.

\section{The gluon self-energies} \label{ii}

The starting point of the computation of the electric and
magnetic self-energies for gluons of color 1
are Eqs.\ (99a) and (99c) of \cite{dhr2f}. For $N_f = 2$ flavors,
\begin{mathletters} \label{Pi11all}
\begin{eqnarray}
\Pi^{00}_{11}(P) & = & - \frac{1}{2}\, g^2  \int 
\frac{{\rm d}^3{\bf k}}{(2 \pi)^3}  \sum_{e_1,e_2 = \pm} (1 + e_1 e_2\,
\hat{\bf k}_1 \cdot \hat{\bf k}_2) \nonumber \\
&  & \times  \left( \frac{1}{p_0 + \epsilon_1 + \epsilon_2 + i \eta}
- \frac{1}{p_0 - \epsilon_1 - \epsilon_2 + i \eta} \right) \,
\frac{ \epsilon_1 \, \epsilon_2 - \xi_1 \, \xi_2 - \phi_1\, \phi_2}{
2 \, \epsilon_1 \, \epsilon_2}  \,\, , \label{Pi1100} \\
\Pi^{t}_{11}(P) & = & - \frac{1}{2}\, g^2  \int 
\frac{{\rm d}^3{\bf k}}{(2 \pi)^3}  \sum_{e_1,e_2 = \pm} 
 (1 - e_1 e_2\,\hat{\bf k}_1 \cdot \hat{\bf p} \;
\hat{\bf k}_2 \cdot \hat{\bf p} ) 
\nonumber \\
&  & \times  \left( \frac{1}{p_0 + \epsilon_1 + \epsilon_2 + i \eta}
- \frac{1}{p_0 - \epsilon_1 - \epsilon_2 + i \eta} \right) \,
\frac{ \epsilon_1 \, \epsilon_2 - \xi_1 \, \xi_2 + \phi_1\, \phi_2}{
2 \, \epsilon_1 \, \epsilon_2}  \,\, . \label{Pi11t}
\end{eqnarray}
\end{mathletters}
Here, ${\bf k}_1 \equiv {\bf k} + {\bf p}/2$ and 
${\bf k}_2 \equiv {\bf k} - {\bf p}/2$,
$\phi_i \equiv \phi^{e_i}_{{\bf k}_i}$ is the gap function for
quasiparticles ($e_i = +1$) or quasi-antiparticles ($e_i = -1$)
with momentum ${\bf k}_i$, the variable $\xi_i$ is defined as
$\xi_i \equiv e_i k_i - \mu$, and
$\epsilon_{i} \equiv \sqrt{\xi_i^2 + \phi_i^2}$ is the
quasiparticle excitation energy. 
The difference between Eqs.\ (\ref{Pi11all})
and Eqs.\ (99) of \cite{dhr2f} is that 
the analytic continuation to real gluon energies $p_0$ has been
made explicit via the $i \eta$ prescription \cite{fetter}, and
that there are no terms $\sim N_i \equiv [\exp(\epsilon_i/T) + 1]^{-1}$, 
since $N_i \equiv 0$ at $T=0$. (Note that $\epsilon_i  \geq \phi_i >0$ in a
superconductor.)

According to Eq.\ (78c) of \cite{dhr2f}, the self-energy
for gluons of color 8 is
\begin{equation} \label{selfen88}
\Pi^{\mu \nu}_{88}(P) = \frac{2}{3}\, \Pi^{\mu \nu}_0(P) 
+ \frac{1}{3}\, \tilde{\Pi}^{\mu \nu}(P) \,\, ,
\end{equation}
where $\Pi^{\mu \nu}_0$ is the standard HDL self-energy. Its
components of relevance for the following are \cite{LeBellac,dhr2f}
\begin{mathletters} \label{Pi0all}
\begin{eqnarray}
\Pi_0^{00} (P) & \simeq & -  3\, m_g^2
\int \frac{{\rm d}\Omega}{4 \pi} 
\left( 1- \frac{p_0}{p_0 + {\bf p} \cdot \hat{\bf k} + i \eta} \right) \,\, ,
\label{Pi000} \\
\Pi_0^{t}(P)  & \simeq &  \frac{3}{2}\, m_g^2
\, \int \frac{{\rm d}\Omega}{4 \pi} \, 
\left[1- (\hat{\bf k} \cdot \hat{\bf p})^2 \right] \,
\frac{p_0}{p_0 + {\bf p} \cdot \hat{\bf k} + i \eta} \,\, ,
\label{Pi0t}
\end{eqnarray}
\end{mathletters}
where $m_g$ is the gluon mass (\ref{gluonmass}).
According to Eqs.\ (102) of \cite{dhr2f}, the electric and magnetic
components of the self-energy $\tilde{\Pi}^{\mu \nu}$ are
\begin{mathletters} \label{Pi88all}
\begin{eqnarray}
\tilde{\Pi}^{00}(P) & = & - \frac{1}{2}\, g^2  \int 
\frac{{\rm d}^3{\bf k}}{(2 \pi)^3}  \sum_{e_1,e_2 = \pm} (1 + e_1 e_2\,
\hat{\bf k}_1 \cdot \hat{\bf k}_2) \nonumber \\
&  & \times  \left( \frac{1}{p_0 + \epsilon_1 + \epsilon_2 + i \eta}
- \frac{1}{p_0 - \epsilon_1 - \epsilon_2 + i \eta} \right) \,
\frac{ \epsilon_1 \, \epsilon_2 - \xi_1 \, \xi_2 + \phi_1\, \phi_2}{
2 \, \epsilon_1 \, \epsilon_2}  \,\, , \label{Pi8800} \\
\tilde{\Pi}^{t}(P) & = & - \frac{1}{2}\, g^2  \int 
\frac{{\rm d}^3{\bf k}}{(2 \pi)^3}  \sum_{e_1,e_2 = \pm} 
(1 - e_1 e_2\,\hat{\bf k}_1 \cdot \hat{\bf p} \; 
\hat{\bf k}_2 \cdot \hat{\bf p} ) 
\nonumber \\
&  & \times  \left( \frac{1}{p_0 + \epsilon_1 + \epsilon_2 + i \eta}
- \frac{1}{p_0 - \epsilon_1 - \epsilon_2 + i \eta} \right) \,
\frac{ \epsilon_1 \, \epsilon_2 - \xi_1 \, \xi_2 - \phi_1\, \phi_2}{
2 \, \epsilon_1 \, \epsilon_2}  \,\, . \label{Pi88t}
\end{eqnarray}
\end{mathletters}
The difference between Eqs.\ (\ref{Pi11all}) and (\ref{Pi88all}) is
the relative sign in front of the term $\sim \phi_1 \, \phi_2$.

In the following, I compute real and imaginary parts of
Eqs.\ (\ref{Pi11all}) and (\ref{Pi88all}) separately, using
the identity
\begin{equation} \label{formula}
\frac{1}{x + i \eta} \equiv {\cal P} \, \frac{1}{x} - i\, \pi \,
\delta(x)
\end{equation}
for the energy denominators in Eqs.\ (\ref{Pi11all}) and
(\ref{Pi88all}),
where ${\cal P}$ stands for the principal value prescription.
I begin with the imaginary parts.

\subsection{Imaginary parts}

In the computation of the imaginary parts it is sufficient 
to consider gluon energies $p_0 \geq 0$, as
${\rm Im}\, \Pi(-p_0,{\bf p}) \equiv 
- {\rm Im}\, \Pi(p_0,{\bf p})$ for all self-energies
(\ref{Pi11all}), (\ref{Pi0all}), (\ref{Pi88all}), which is easy
to prove using Eq.\ (\ref{formula}), and noting that $\delta(x) = \delta(-x)$.
It is instructive to first consider the imaginary parts of the
HDL self-energies. For electric and magnetic
gluons one obtains the
standard result \cite{LeBellac,RDPphysica}
\begin{mathletters} \label{ImPi0}
\begin{eqnarray}
{\rm Im}\, \Pi^{00}_0(P) & \simeq & - \pi \; \frac{3}{2}\, m_g^2 \,
\frac{p_0}{p}\, \theta(p - p_0)\,\, , \label{ImPi000} \\
{\rm Im}\, \Pi^t_0(P) & \simeq & - \pi \; \frac{3}{4}\, m_g^2 \,
\frac{p_0}{p} \, \left( 1 - \frac{p_0^2}{p^2} \right) \,
\theta(p-p_0)\,\, , \label{ImPi0t}
\end{eqnarray}
\end{mathletters}
corresponding to Landau damping for space-like gluons.

For ${\rm Im}\, \Pi^{00}_{11}(P)$ and $p_0 >0$,
the delta-function originating from the
first energy denominator in Eq.\ (\ref{Pi1100}) has no
support, since $\epsilon_i \geq \phi_i >0$.
For the delta-function originating from the second
energy denominator, use the fact that the interesting range
of gluon energies in the gap equation (\ref{gapequation}) is
$p_0 \ll \mu$, as the quarks in the Cooper pair are close to the
Fermi surface where $\epsilon_i \sim \phi_i \ll \mu$.
Then, the only term in the sums over $e_1,\, e_2$ in Eq.\
(\ref{Pi1100}) that one has to keep
is the one where $e_1 = e_2 = +1$; for $e_i = -1$,  
$\epsilon_i \simeq |k_i + \mu| \sim \mu $ is far from the Fermi surface and
the delta-function has no support for $p_0 \ll \mu$.

Furthermore, the fermion momentum ${\bf k}$ in the loop is close to the
Fermi surface, $k \sim \mu$, while the interesting range of gluon momenta 
in the gap equation is $p \ll k$, which allows to approximate
\begin{mathletters} \label{approx1}
\begin{eqnarray}
\hat{\bf k}_1 \cdot \hat{\bf k}_2 & \simeq  & 1   \,\, , \\
\xi_1 & \simeq & \xi + \frac{{\bf p} \cdot \hat{\bf k}}{2} 
\equiv \xi_+ \,\, , \\
\xi_2 & \simeq & \xi - \frac{{\bf p} \cdot \hat{\bf k}}{2}
\equiv \xi_-\,\, ,
\end{eqnarray}
\end{mathletters}
where $\xi \equiv k - \mu$.
Finally, setting $\phi_1 \simeq \phi_2 \equiv \phi$, 
and defining $\epsilon_\pm \equiv \sqrt{\xi_\pm^2 + \phi^2}$, one obtains
\begin{equation}
{\rm Im}\, \Pi^{00}_{11} (P) \simeq
- \pi\, g^2 \int \frac{{\rm d}^3 {\bf k}}{(2 \pi)^3} \;
\delta(p_0 - \epsilon_+ - \epsilon_-) \;
\frac{ \epsilon_+ \, \epsilon_- - \xi_+ \, \xi_- - \phi^2}{
2 \, \epsilon_+ \, \epsilon_- }  \,\, . \label{ImPi1100}
\end{equation}
First, note that since $\epsilon_+ + \epsilon_- \geq 2 \phi$,
the delta-function has no support for $p_0 < 2 \phi$:
the imaginary part of the gluon self-energy vanishes below
the threshold $2 \phi$ for quasiparticle-quasihole excitations.
This is clearly different from the HDL self-energy which
has an imaginary part for {\em all\/} gluon energies
$0<p_0\leq p$, cf.\ Eq.\ (\ref{ImPi000}).

\begin{figure}
\hspace*{2cm}
\mbox{\epsfig{figure=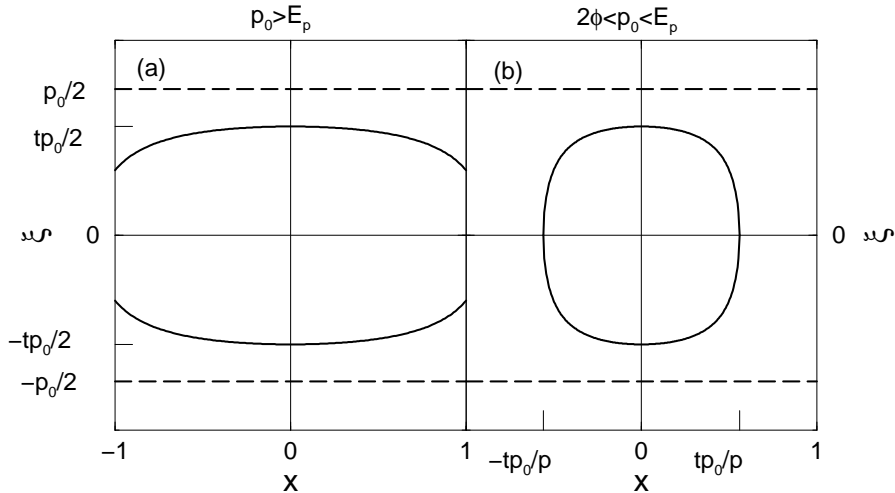,height=12cm,angle=270}}
\caption{Solid lines: the function $\xi^*(x)$ from Eq.\ (\ref{fix}) for
(a) $p_0 > E_p$, and (b) $2 \phi \leq p_0 \leq E_p$.
Dashed lines represent the case $\phi = 0$.}
\label{fig1}
\end{figure}

The integration over ${\bf k}$ is best done in spherical coordinates,
choosing ${\bf p} = (0,0,p)$. Then, the integration over the polar angle 
$\varphi$
becomes trivial. Replacing the integration over azimuthal angle $\theta$
by $x \equiv \cos \theta$ and over $k$ by $\xi \equiv k - \mu$,
one can use the delta-function to either
perform the integration over $x$, or over $\xi$.
Choosing the latter, for a given value of $x$
one finds that there are
two roots of the argument of
the delta-function, 
which differ by an overall sign,
\begin{equation} \label{fix}
\xi^*(x) = \pm \frac{p_0}{2}\;  
\sqrt{ 1 - \frac{4 \, \phi^2}{p_0^2 - p^2\, x^2}}\,\,.
\end{equation}
It is instructive to draw these solutions in the $(x,\xi)$-plane,
cf.\ Fig.\ \ref{fig1}. In part (a) of this figure, corresponding
to $p_0 > E_p$, where
\begin{equation} \label{Ep}
E_p \equiv \sqrt{p^2 + 4\, \phi^2}\,\, ,
\end{equation}  
these two roots
exist for all values of $x \in [-1,1]$, while in part (b), corresponding
to $2 \phi \leq p_0 \leq E_p$, they merge at $\xi=0$ and 
$x = \pm t\,p_0/p$,
where
\begin{equation} \label{t}
t \equiv \sqrt{1- \frac{4 \, \phi^2}{p_0^2}}\,\,;
\end{equation} 
there is no solution for $1 \geq |x| > t\, p_0/p$.
For $p_0 < 2 \phi$, there is no solution either, as already mentioned
above. Note that $E_p = 2 \phi$ for $p = 0$, while $E_p \simeq p$
for $p \gg 2 \phi$. Analogously,
$t = 0$ for $p_0 = 2 \phi$, while $t \simeq 1$
for $p_0 \gg 2 \phi$.

In the limit $\phi \rightarrow 0$, the $x$-integration 
decouples from the integration over $\xi$, and by Eq.\ (\ref{fix}),
the value of the latter is simply constant, $\xi^* \equiv \pm p_0/2$.
This is indicated by the dashed lines in Fig.\ \ref{fig1}.

Evaluating the $\xi$-integral at the roots (\ref{fix}), one
obtains
\begin{equation}
{\rm Im}\, \Pi^{00}_{11} (P) \simeq - \pi \; 
\frac{3}{2}\, m_g^2 \; \theta(p_0 - 2 \, \phi) \; \frac{4\, \phi^2}{p_0\, p}
\int_0^u {\rm d}y \, \frac{y^2}{(1-y^2)^{3/2} \sqrt{t^2 - y^2}}
\,\, ,
\end{equation}
where $u \equiv {\rm min}\, (t,p/p_0)$. The integral
can be expressed in terms of elliptic integrals, yielding the
final answer
\begin{eqnarray}
{\rm Im}\, \Pi^{00}_{11} (P) & \simeq & - \pi \; 
\frac{3}{2}\, m_g^2 \; \theta(p_0 - 2 \, \phi) \; \frac{p_0}{p}
\nonumber \\
& \times & \left\{ \theta(E_p - p_0)\, 
\left[ {\bf E}(t) - s^2 {\bf K}(t) \right] + 
\theta(p_0 - E_p) \, \left[  E(\alpha,t) - s^2 F(\alpha,t) - \frac{p}{p_0}
\, \sqrt{1- \frac{4 \, \phi^2}{p_0^2 - p^2}} \right] \right\}
\,\, , \label{ImPi1100final}
\end{eqnarray}
where $s \equiv \sqrt{1-t^2} = 2 \, \phi/p_0$,
$\alpha \equiv \arcsin [p/(t\, p_0)]$, and $F(\alpha,t)$ and $E(\alpha,t)$
are elliptic integrals of the first and second kind,
\begin{equation}
F(\alpha,t) \equiv \int_0^\alpha {\rm d} x \; 
\frac{1}{\sqrt{1-t^2 \, \sin^2 x}}
\;\;\;\; , \;\;\;\;
E(\alpha,t) \equiv \int_0^\alpha {\rm d} x \; \sqrt{1-t^2 \, \sin^2 x}\,\, ,
\end{equation}
while ${\bf K}(t) \equiv F(\pi/2,t)$ and ${\bf E}(t) \equiv E(\pi/2,t)$
are {\em complete\/} elliptic integrals of the first and second kind.
At $p_0 = E_p$, ${\rm Im}\, \Pi^{11}_{00}$ is continuous, since then
$\alpha = \pi/2$ and the square root in Eq.\ (\ref{ImPi1100final})
vanishes.

It is important to note that the limit $\phi \rightarrow 0$ of the
result (\ref{ImPi1100final}) exists for all $P$: as $\phi \rightarrow 0$,
$\alpha \rightarrow \arcsin (p/p_0)$ and $t \rightarrow 1$,
such that $E(\alpha,t) \rightarrow p/p_0$, while
$F(\alpha,t) \rightarrow \frac{1}{2} \, \ln [(p_0+p)/(p_0-p)]$.
Therefore, for $p_0 > E_p$, the imaginary part
vanishes like $s^2 \sim \phi^2/p_0^2$. In the other region, $p_0 \leq E_p$,
when $t \rightarrow 1$, ${\bf E}(t) \rightarrow 1$, while
$s^2 {\bf K}(t) \rightarrow 0$. In summary,
the $\phi \rightarrow 0$ limit of ${\rm Im}\, \Pi^{00}_{11}(P)$
is the imaginary part (\ref{ImPi000}) of the electric HDL self-energy,
\begin{equation}
\lim_{\phi \rightarrow 0} \; {\rm Im}\, \Pi^{00}_{11}(P) \equiv
{\rm Im}\, \Pi^{00}_0 (P)\,\, .
\end{equation}

\begin{figure}
\hspace*{1.5cm}
\mbox{\epsfig{figure=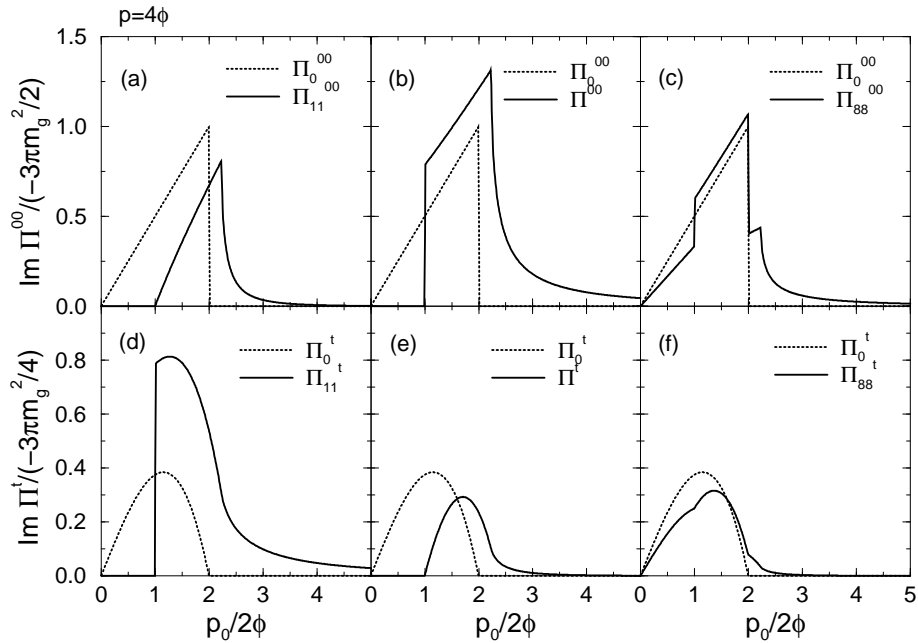,height=12cm,angle=270}}
\caption{The imaginary parts of the gluon self-energies
as function of energy $p_0$ for $p = 4\, \phi$. Figures (a), (b), and
(c) show the electric self-energies, (d), (e), and (f) the
magnetic self-energies. The solid lines in (a) and (d) 
are the imaginary parts of the self-energies for gluons of color 1, in
(b) and (e) they are the imaginary parts of the functions $\tilde{\Pi}^{00}$
and $\tilde{\Pi^t}$ occurring in the self-energies for gluons of color 8,
the imaginary parts of which are shown in (c) and (f).
The dashed lines in all graphs are the imaginary parts of the
HDL self-energies.}
\label{fig2}
\end{figure}

The computation of the imaginary part of the magnetic self-energy 
for gluons of color 1 is completely analogous. With the approximation
\begin{equation}
\hat{\bf k}_1 \cdot \hat{\bf p}\; \hat{\bf k}_2 \cdot \hat{\bf p}
\simeq ( \hat{\bf k} \cdot \hat{\bf p} )^2 \,\, ,
\end{equation}
valid when $p \ll k \sim \mu$, one derives
\begin{eqnarray}
{\rm Im}\, \Pi^{t}_{11} (P) & \simeq & - \pi \; 
\frac{3}{4}\, m_g^2 \; \theta(p_0 - 2 \, \phi) \; \frac{p_0}{p}
\; \left\{ \theta(E_p - p_0)\, 
\left[ \left( 1 - \frac{p_0^2}{p^2} \right)\, {\bf E}(t) + s^2\,
\frac{p_0^2}{p^2}\, {\bf K}(t) \right] \right. \nonumber \\
&    & \left. + \;
\theta(p_0 - E_p) \, \left[  
\left( 1 - \frac{p_0^2}{p^2} \right) \left( E(\alpha,t) - \frac{p}{p_0}
\, \sqrt{1- \frac{4 \, \phi^2}{p_0^2 - p^2}}\right) + s^2\,
\frac{p_0^2}{p^2}\,F(\alpha,t)  \right] \right\}
\,\, . \label{ImPi11tfinal}
\end{eqnarray}
Again, ${\rm Im}\, \Pi^t_{11}$ is continuous at $p_0 = E_p$, and
the limit $\phi \rightarrow 0$ exists for all $P$,
\begin{equation}
\lim_{\phi \rightarrow 0} \; {\rm Im}\, \Pi^{t}_{11}(P) \equiv
{\rm Im}\, \Pi^t_0 (P)\,\, .
\end{equation}
The computation of the imaginary parts of the
electric and magnetic self-energies (\ref{Pi88all}) is
straightforward; the final result is
\begin{eqnarray}
{\rm Im}\, \tilde{\Pi}^{00} (P) & \simeq & - \pi \, 
\frac{3}{2}\, m_g^2 \, \theta(p_0 - 2 \, \phi) \, \frac{p_0}{p}
\, \left\{ \theta(E_p - p_0)\, {\bf E}(t) + 
\theta(p_0 - E_p) \, \left[  E(\alpha,t)  - \frac{p}{p_0}
\, \sqrt{1- \frac{4 \, \phi^2}{p_0^2 - p^2}} \right] \right\}
\,\, , \label{ImtildePi00final} \\
{\rm Im}\, \tilde{\Pi}^{t} (P) & \simeq & - \pi \,
\frac{3}{4}\, m_g^2 \, \theta(p_0 - 2 \, \phi) \, \frac{p_0}{p}
\, \left\{ \theta(E_p - p_0)\, 
\left[ \left( 1 - \frac{p_0^2}{p^2}\, (1+s^2) \right)\, {\bf E}(t) - s^2\,
\left(1- 2\, \frac{p_0^2}{p^2} \right)\, {\bf K}(t) \right] \right. 
\nonumber \\
&    &  + \;
\theta(p_0 - E_p) \, \left[  
\left( 1 - \frac{p_0^2}{p^2}\,(1+s^2) \right) E(\alpha,t) 
- \left( 1 - \frac{p_0^2}{p^2} \right) \frac{p}{p_0}
\, \sqrt{1- \frac{4 \, \phi^2}{p_0^2 - p^2}} \right. \nonumber \\
&    & \left. \left. \hspace*{2.2cm} - \; s^2
\left( 1 - 2 \, \frac{p_0^2}{p^2} \right) F(\alpha,t)  \right] \right\}
\,\, . \label{ImtildePitfinal}
\end{eqnarray}
As in the previous cases, these functions are continuous at
$p_0 = E_p$, and the limit $\phi \rightarrow 0$ exists;
together with Eq.\ (\ref{selfen88}) one concludes
\begin{eqnarray}
\lim_{\phi \rightarrow 0} \; {\rm Im}\, {\Pi}^{00}_{88} (P) & \equiv &
{\rm Im}\, \Pi^{00}_0 (P)\,\, , \\
\lim_{\phi \rightarrow 0} \; {\rm Im}\, {\Pi}^{t}_{88} (P) & \equiv &
{\rm Im}\, \Pi^{t}_0 (P)\,\, .
\end{eqnarray}
Figure \ref{fig2} shows the imaginary parts 
for a representative value of the gluon momentum $p = 4\, \phi$ as
functions of gluon energy $p_0$ (in units of $2\, \phi$).
Comparing the self-energies in a color superconductor (solid lines)
with those in a normal conductor (dashed lines) one observes the following
two main differences. First, as is obvious from Eqs.\
(\ref{ImPi1100final}), (\ref{ImPi11tfinal}),
(\ref{ImtildePi00final}), and (\ref{ImtildePitfinal}),
${\rm Im} \Pi^{00,t}_{11}$ and ${\rm Im} \tilde{\Pi}^{00,t}$
vanish below $p_0 = 2\, \phi$,
representing the fact that for such small gluon energies
it is impossible to excite quasiparticle-quasihole pairs which would lead to
non-vanishing self-energies. For the eighth gluon, however, the functions
${\rm Im} \Pi^{00,t}_{88}$ do not vanish because of the admixture
of the HDL self-energy, Eq.\ (\ref{selfen88}).
Second, while the imaginary parts of the HDL self-energies vanish
above the light cone $p_0 = p$, the self-energies in a superconductor
are never zero, although they rapidly fall off above $p_0 = E_p$.
As will be discussed in more detail in Sec.\ \ref{iii}, this indicates
that, already at one-loop order,
gluons are damped on the quasi-particle mass shell \cite{carterreddy}.
Both features persist for all gluon momenta $p$. 

Another interesting property is that the electric self-energy for
gluons of color 1 and the magnetic self-energy for gluons of color 8 are
continuous, while the magnetic self-energy for gluons of color 1
has a discontinuity at $p_0 = 2\, \phi$
and the electric self-energy for gluons of color 8 has one
at $p_0 = 2\, \phi$ and one at $p_0 = p$, the latter arising from
the discontinuity of the HDL part in Eq.\ (\ref{selfen88}).
This leads to characteristic features in the real parts, as
will be discussed below.
Again, these features are generic for all gluon momenta $p$.

I conclude the discussion noting that for $p_0 \gg \phi$ the
imaginary parts of the gluon self-energies approach the
corresponding HDL expressions. They deviate from the
latter only for $p_0 \sim \phi$.

\subsection{Real parts}

There are in principle two ways to compute the real parts of the
gluon self-energies, either as a principal value integral
from Eq.\ (\ref{formula}), or from the dispersion integral
\begin{equation} \label{dispersionintegral}
{\rm Re}\, \Pi(p_0,{\bf p}) \equiv
\frac{1}{\pi}\, {\cal P} \int_0^\infty {\rm d} \omega \;
{\rm Im}\, \Pi (\omega,{\bf p})\, \left( \frac{1}{\omega + p_0}
+ \frac{1}{\omega - p_0} \right) +  C \,\, ,
\end{equation}
where $C$ is a (subtraction) constant, and where
$\Pi(p_0,{\bf p})$ is assumed to be analytic in the upper
complex $p_0$-plane.

In the HDL case, Eq.\ (\ref{dispersionintegral}) gives
for electric gluons
\begin{mathletters} \label{realHDL}
\begin{equation}
{\rm Re}\, \Pi^{00}_0(p_0,{\bf p}) \simeq - 3\,m_g^2
\left( 1 - \frac{p_0}{2 p} \, \ln \left| \frac{p_0 + p}{p_0 - p}
\right| \right) + C^{00}_0 \,\, ,
\end{equation}
and for magnetic gluons 
\begin{equation}
{\rm Re}\, \Pi^t_0(p_0,{\bf p}) \simeq  \frac{3}{2}\, m_g^2
\left[ \frac{p_0^2}{p^2} + \left(1-\frac{p_0^2}{p^2}\right) 
\frac{p_0}{2 p}\, \ln \left| \frac{p_0 + p}{p_0 - p} \right|  
 - \frac{2}{3}\right] + C^t_0\,\, . \label{RePi0t}
\end{equation}
\end{mathletters}
Comparison with a direct calculation via Eq.\ (\ref{Pi0all})
determines 
\begin{equation} \label{constants}
C^{00}_0 \equiv 0 \;\;\;\; , \;\;\;\;\; 
C^t_0 \equiv m_g^2\,\, ,
\end{equation}
such that the term $-2/3$ in Eq.\ (\ref{RePi0t}) is cancelled by
$C^t_0$ and the magnetic self-energy has the correct zero-energy
limit, ${\rm Re}\, \Pi^t_0(0,{\bf p}) = 0$, representing
the absence of magnetic screening to one-loop order. 
Thus, the standard expressions for the real parts of the HDL
self-energies are recovered \cite{LeBellac,RDPphysica}.

It should be remarked that the constant $C^t_0$ is not a
subtraction constant in the mathematical sense. A true subtraction
constant would be required if ${\rm Im}\, \Pi^t_0(p_0,{\bf p})$ did
not vanish when $p_0 \rightarrow \infty$. However, as can
be seen from Eq.\ (\ref{ImPi0}), the imaginary parts
of both electric and magnetic HDL self-energy are zero in this limit.
The origin of a nonzero constant $C^t_0$ is purely physical. As discussed
in the last subsection, within the present approximations
only quasiparticle-quasihole excitations
near the Fermi surface give rise to the imaginary part of $\Pi^t_{aa}$. 
This holds also in the normal conducting phase, {\it i.e.}, for
${\rm Im}\, \Pi^t_0$.
When computing the real part
via the dispersion integral (\ref{dispersionintegral})
these contributions are correctly taken into account. However,
what is not accounted for is that, in contrast to the
electric self-energies,
the real parts of the magnetic self-energies also receive a contribution
from quasiparticle-antiparticle excitations, cf.\ the discussion
in Refs.\ \cite{dhr2f,dhr3f}. There,
I explicitly computed this contribution 
for $p_0 = p = 0$ in a color superconductor, 
and found it to be $m_g^2$, see also \cite{sonstephanov,hong}.
As antiparticles are always far from the
Fermi surface, it actually does not matter whether one considers these
excitations in a superconductor or in a normal conductor,
or for nonzero gluon energy and momentum, provided 
$p_0,\, p \ll \mu$, as is the case here.
One therefore concludes that the constant $C^t_0 = m_g^2$ arises precisely
from particle-antiparticle excitations, and that $C^{00}_0 = 0$,
because the electric self-energies do not receive such contributions.

For the real parts of the
gluon self-energies in a two-flavor color superconductor,
I was not able to find expressions in closed
form, so that the computation had to be done numerically.
One can either directly use Eqs.\ (\ref{Pi11all}), 
(\ref{Pi88all}) with Eq.\ (\ref{formula}),
or the dispersion integral (\ref{dispersionintegral}).
It turns out that the second way is numerically far simpler;
details can be found in Appendix \ref{app2}.
This method does not specify the values of the
subtraction constants $C^{00}_{aa}$ 
and $C^t_{aa}$, $a=1,8$, but from 
the above discussion on the origin of the result (\ref{constants}),
one immediately concludes
\begin{equation}
C^{00}_{11} \equiv C^{00}_{88} \equiv C^{00}_0 = 0
\;\;\;\; , \;\;\;\;\;
C^{t}_{11} \equiv C^{t}_{88} \equiv C^{t}_0 = m_g^2\,\,.
\end{equation}
This has also been confirmed by 
a direct numerical calculation of the real parts via
Eqs.\ (\ref{Pi11all}), (\ref{Pi88all}) with Eq.\ (\ref{formula}).

\begin{figure}
\hspace*{1.5cm}
\mbox{\epsfig{figure=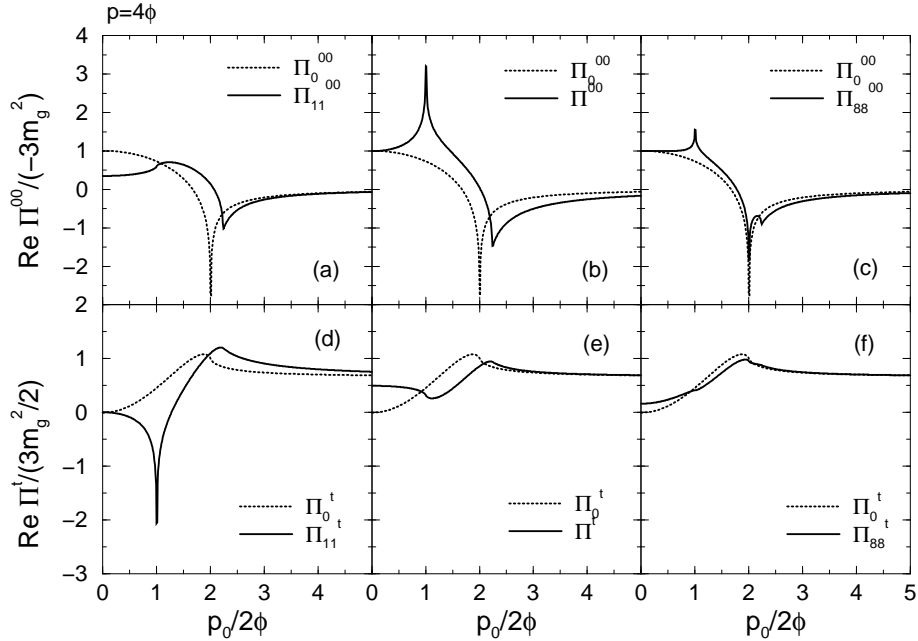,height=12cm,angle=270}}
\caption{Same as in Fig.\ \ref{fig2}, but for the real parts of the 
gluon self-energies.}
\label{fig3}
\end{figure}

Figure \ref{fig3} shows the real parts of the gluon self-energies
corresponding to the imaginary parts of Fig.\ \ref{fig2}.
The various discontinuities of the imaginary parts discussed above
appear in the real parts as logarithmic 
singularities, due to the relationship between
real and imaginary parts dictated by the dispersion integral 
(\ref{dispersionintegral}).
Consequently, ${\rm Re}\, \Pi^{00}_0$ has a logarithmic singularity
at $p_0 = p$, cf.\ Figs.\ \ref{fig3} (a), (b), (c), 
on account of the corresponding
discontinuity of the imaginary part, while ${\rm Re}\, \Pi^t_0$ is
regular, see Figs.\ \ref{fig3} (d), (e), (f). 

By the same line of arguments ${\rm Re}\, \Pi^{00}_{11}$, shown
in Fig.\ \ref{fig3} (a), is regular.
The peak at $p_0 \simeq E_p$ visible in this figure
is caused by the rapid decrease of the
imaginary part in Fig.\ \ref{fig2} (a), but it is not a true
singularity as long as $\phi$ is nonzero. It becomes a
singularity (the same as the one of ${\rm Re}\, \Pi^{00}_0$) in the
limit $\phi \rightarrow 0$.
In contrast, ${\rm Re}\, \Pi^t_{11}$ has a true logarithmic singularity
at $p_0 = 2\, \phi$, Fig.\ \ref{fig3} (d). This singularity persists
for all nonzero $\phi$, but obviously moves towards $p_0 = 0$ as
$\phi \rightarrow 0$.

As seen in Fig.\ \ref{fig3} (b), 
the discontinuity of ${\rm Im}\, \tilde{\Pi}^{00}$ at
$p_0 = 2\, \phi$ causes a singularity in ${\rm Re}\, \tilde{\Pi}^{00}$.
Together with the singularity of ${\rm Re}\, \Pi^{00}_0$ at $p_0 = p$,
this causes two logarithmic singularities in ${\rm Re}\, \Pi^{00}_{88}$,
cf.\ Fig.\ \ref{fig3} (c).
As for ${\rm Re}\, \Pi^{00}_{11}$, 
the peak at $p_0 = E_p$ is not a singularity, but is due to the
sharp (but continuous) decrease of ${\rm Im}\, \tilde{\Pi}^{00}$.
Only in the limit $\phi \rightarrow 0$, this becomes a
true singularity and merges with the one of ${\rm Re}\, \Pi^{00}_0$
at $p_0 = p$. The singularity at $p_0 = 2\, \phi$ persists for all
nonzero $\phi$.
Finally,
${\rm Re}\, \tilde{\Pi}^t$ is regular, cf.\ Fig.\ \ref{fig3} (e). From
Eq.\ (\ref{selfen88}) and the regularity of ${\rm Re}\,
\Pi^t_0$ one then concludes that the same holds for 
${\rm Re}\, \Pi^t_{88}$, Fig.\ \ref{fig3} (f).  

{}From the above discussion one concludes that 
\begin{eqnarray}
\lim_{\phi \rightarrow 0} \; {\rm Re}\, \Pi^{00}_{11}(P) & \equiv &
{\rm Re}\, \Pi^{00}_0 (P)\,\, , \\
\lim_{\phi \rightarrow 0} \; {\rm Re}\, \Pi^t_{88}(P) & \equiv &
{\rm Re}\, \Pi^t_0 (P)\,\, .
\end{eqnarray}
These equations hold for all $P$ (even for the first equation, 
if, for $p_0 = p$, one includes $\infty$ as a possible value
for $\lim {\rm Re}\, \Pi^{00}_{11}$).

The functions ${\rm Re}\, \Pi^t_{11}$ and ${\rm Re}\, \Pi^{00}_{88}$
also converge towards the corresponding HDL expressions, except
in a region of size $\sim \phi$
near the logarithmic singularity at $p_0 = 2\, \phi$, where
the deviation from the HDL self-energies is large 
(infinite at $p_0 = 2\, \phi$).
Of course, in the limit $\phi \rightarrow 0$, the
point $p_0 = 2\, \phi$ moves towards $p_0 = 0$, and the size of
the region where there are substantial deviations shrinks to zero.
In this sense, the limit $\phi \rightarrow 0$ 
for ${\rm Re}\, \Pi^t_{11}$ and ${\rm Re}\, \Pi^{00}_{88}$ still exists
for all $P$ and coincides with the corresponding HDL expressions.

As for the imaginary parts, the real parts of the gluon
self-energies in a color superconductor approach the corresponding 
HDL expressions when $p_0 \gg \phi$. They deviate from the latter
only for $p_0 \sim \phi$.

\newpage
\section{The spectral densities} \label{iii}

The spectral densities are defined by the relations
\begin{equation}\label{specdens}
\rho^{00,t}(p_0,{\bf p}) \equiv \frac{1}{\pi} \, 
{\rm Im}\, \Delta^{00,t}(p_0 + i \eta,{\bf p})\,\, .
\end{equation}
When ${\rm Im}\, \Pi^{00,t}(p_0,{\bf p}) \neq 0$, the spectral densities
are regular and Eq.\ (\ref{specdens}) is identical to
\begin{mathletters} \label{regular}
\begin{eqnarray} 
\rho^{00}(p_0,{\bf p}) & = & \frac{1}{\pi} \, 
\frac{{\rm Im}\, \Pi^{00}(p_0,{\bf p})}{
\left[p^2 - {\rm Re}\, \Pi^{00}(p_0,{\bf p}) \right]^2
+ \left[ {\rm Im}\, \Pi^{00}(p_0,{\bf p}) \right]^2} 
\, , \\
\rho^{t}(p_0,{\bf p}) & = & \frac{1}{\pi} \, 
\frac{{\rm Im}\, \Pi^{t}(p_0,{\bf p})}{
\left[p_0^2 - p^2 - {\rm Re}\, \Pi^{t}(p_0,{\bf p}) \right]^2
+ \left[ {\rm Im}\, \Pi^{t}(p_0,{\bf p}) \right]^2} 
\,\, .
\end{eqnarray}
\end{mathletters}
When ${\rm Im}\, \Pi^{00,t}(p_0,{\bf p}) = 0$, for given
gluon momentum ${\bf p}$ the spectral densities
have simple poles, determined
by the solution of 
\begin{mathletters} \label{disp}
\begin{equation} \label{disp00}
\left[ p^2 - {\rm Re}\, \Pi^{00}(p_0,{\bf p}) 
\right]_{p_0 = \omega^{00}({\bf p})} = 0
\end{equation}
in the electric case and
\begin{equation} \label{dispt}
\left[ p_0^2 - p^2 - {\rm Re}\, \Pi^t(p_0,{\bf p}) 
\right]_{p_0 = \omega^t({\bf p})} = 0
\end{equation}
\end{mathletters}
in the magnetic case,
and are zero elsewhere, {\it i.e.},
\begin{equation} \label{singular}
\rho^{00,t}(p_0,{\bf p})  =  - Z^{00,t}({\bf p}) \left\{ \frac{}{}
\delta \left[ p_0 - \omega^{00,t}({\bf p})\right] 
- \delta \left[p_0 + \omega^{00,t}({\bf p}) \right] \;
 \right\} \,\, .
\end{equation}
The solutions $\omega^{00,t}({\bf p})$ to Eqs.\ (\ref{disp})
define the quasiparticle dispersion relations, and
\begin{equation} 
Z^{00,t}({\bf p}) \equiv  
\left( \left| \frac{\partial (\Delta^{00,t})^{-1}(p_0,{\bf p})}{
\partial p_0} \right| \right)^{-1}_{p_0 = \omega^{00,t}({\bf p})}
\end{equation}
are the residues at the poles $p_0 = \omega^{00,t}({\bf p})$.

\begin{figure}
\hspace*{3.5cm}
\mbox{\epsfig{figure=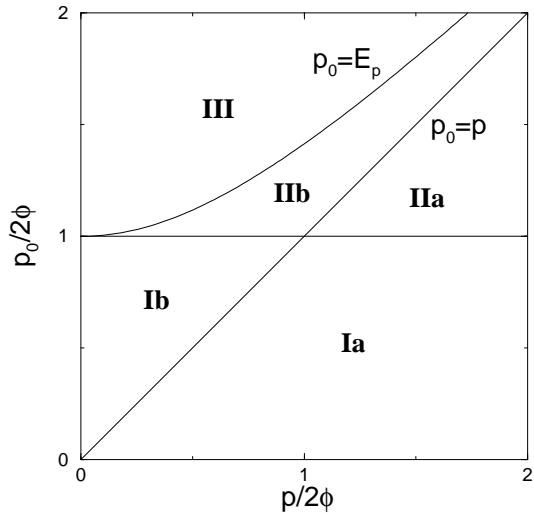,height=7cm,angle=270}}
\caption{Schematic plot of the $(p_0, p)$-plane. The HDL
spectral density assumes the form (\ref{regular}) in regions Ia and
IIa, and the form (\ref{singular}) in regions Ib, IIb, and III. 
In a two-flavor color superconductor, the spectral density is always
of the form (\ref{regular}) in regions IIa, IIb, and III, and
of the form (\ref{singular}) in regions Ia and Ib.}
\label{fig4}
\end{figure} 

\begin{figure}
\hspace*{2cm}
\mbox{\epsfig{figure=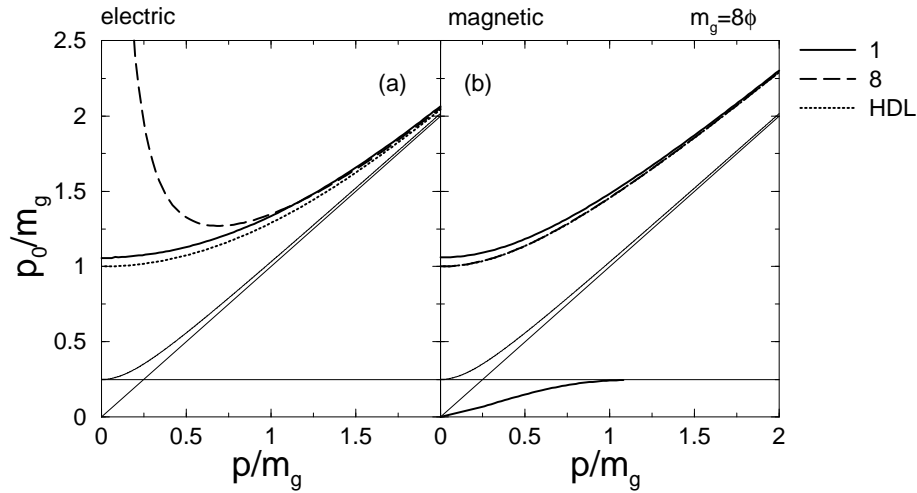,height=12cm,angle=270}}
\caption{Dispersion relations for (a) electric and (b) magnetic
gluons, $m_g = 8 \, \phi$. 
The different regions of Fig.\ \ref{fig4} are indicated by thin
solid lines. The dotted curves are the HDL dispersion relations.
The solid curves are for gluons of color 1, the dashed curves for
gluons of color 8. For magnetic gluons, the dispersion curve
for gluons of color 8 and the HDL dispersion curve are visually
indistinguishable. Note the additional dispersion branch below 
$p_0 = 2 \, \phi$ (region Ia of Fig.\ \ref{fig4}) for
magnetic gluons of color 1. This branch merges
with the continuum for $p_0 \geq 2\, \phi$ above $p \simeq 1.1\, m_g$. }
\label{fig5}
\end{figure}

\begin{figure}
\vspace*{-1cm}
\hspace*{1.5cm}
\mbox{\epsfig{figure=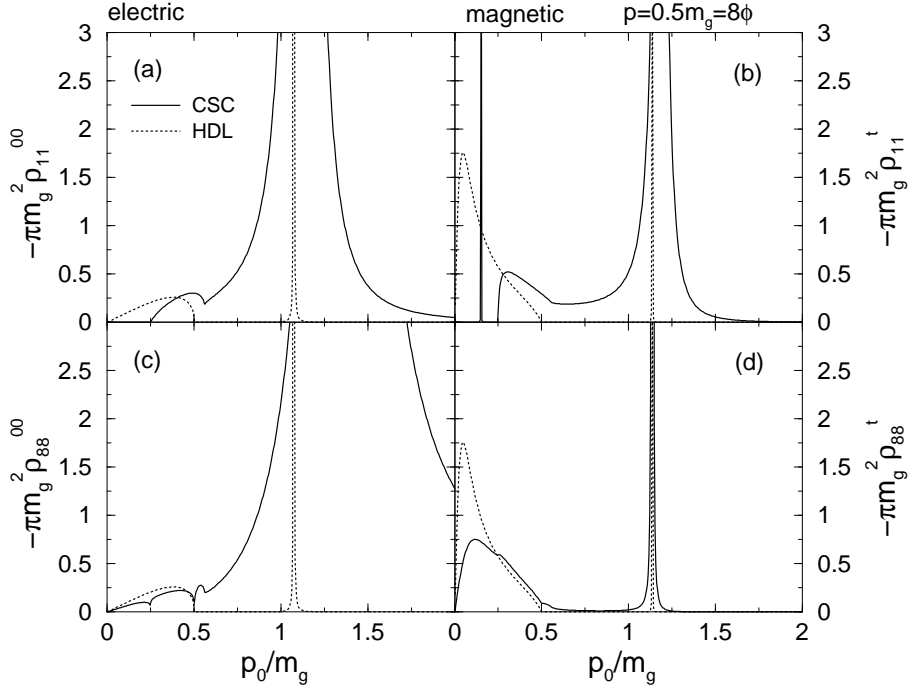,height=12cm,angle=270}}
\caption{Electric (a,c) and magnetic (b,d)
spectral densities for gluons of color 1
(a,b) and color 8 (c,d), for a fixed value of
the gluon momentum, $p = m_g/2$, and $m_g = 8 \, \phi$.
The dotted curves represent the HDL spectral densities. In order to make
the delta-function at the quasiparticle dispersion curves visible, 
I used the regular form (\ref{regular}) of the spectral
density with a numerically small, but nonzero imaginary part.}
\label{fig6}
\end{figure}

In order to understand the structure of the gluon spectral densities
in a two-flavor color superconductor, it is instructive to first
remember the structure of the HDL spectral densities in
the $(p_0, p)$-plane, cf.\ Fig.\ \ref{fig4}. 
Below the light cone $p_0 = p$ (regions Ia and IIa in
Fig.\ \ref{fig4}), the imaginary parts of
both electric and magnetic self-energies are nonzero, {\it i.e.},
gluons are Landau-damped. Therefore, the spectral density is regular 
and of the form (\ref{regular}). Above the light cone
(regions Ib, IIb, and III), the HDL spectral
density is of the form (\ref{singular}), with the dispersion
relations $p_0 = \omega^{00}_0({\bf p})$ for electric and 
$p_0 = \omega^t_0({\bf p})$ for magnetic gluons. The explicit
formulae for the electric ({\it i.e.}\ longitudinal) and 
magnetic ({\it i.e.}\ transverse) dispersion relations can
be found in \cite{LeBellac,RDPphysica}; they are
shown graphically in Fig.\ \ref{fig5} (dotted curves).
In weak coupling, $m_g \gg \phi$, and since
$\omega^{00}_0(0) = \omega^t_0(0) = m_g$, 
the electric and magnetic dispersion curves 
lie entirely in region III of Fig.\ \ref{fig4};
in regions Ib and IIb the HDL spectral densities vanish.
Figure \ref{fig6} further clarifies the
shape of the HDL spectral densities as functions of $p_0$
for an exemplary value of $p$.

On the other hand, in a two-flavor color superconductor,
the imaginary parts of the self-energies never vanish 
for energies $p_0 \geq 2\, \phi$, {\it i.e.}, in regions IIa, IIb, and
III of Fig.\ \ref{fig4}. Thus, in these regions
the spectral densities are {\em always\/} regular and of
the form (\ref{regular}). Nevertheless, for $p_0 \geq E_p$
(region III) the imaginary parts are small, ${\rm Im}\,\Pi(p_0,{\bf p})
\sim  \phi^2$. One can still define
dispersion relations $p_0 = \omega({\bf p})$ through Eqs.\ (\ref{disp}).
These are shown in Fig.\ \ref{fig5}.
In weak coupling, $m_g \gg \phi$, 
these dispersion curves are nearly identical to the corresponding
HDL dispersion curves, and consequently also lie
in region III. 
(A peculiar feature is the dispersion curve for electric gluons of
color 8, which has a negative slope for small momenta $p \sim \phi$.
At present, I cannot exclude that this behavior is merely
an artifact caused by neglecting the mixing of
these gluons with the excitations of the condensate.
This will be clarified in 
a future publication \cite{shovrisch}.)

Although the quasiparticles are not stable, like in the HDL case, 
they decay on a comparatively
long time scale $\Gamma^{-1} \sim [{\rm Im}\, \Pi(p_0,{\bf p})/p_0]^{-1}
\sim p_0/\phi^2$.
The spectral densities resemble ``smeared'' delta-functions which peak
at $p_0 = \omega({\bf p})$ 
and have a width $\sim {\rm Im}\, \Pi(p_0,{\bf p})
\sim \phi^2$. 
This can also be seen in Fig.\ \ref{fig6}, where
the electric and magnetic spectral densities in a color superconductor
are shown as functions of $p_0$ for a fixed value of the
gluon momentum, $p = m_g/2$.

For $p_0 < 2\, \phi$ (regions Ia, Ib in Fig.\ \ref{fig4}) the
imaginary parts of the gluon self-energies in a color superconductor
vanish and the spectral densities are of the form (\ref{singular}).
It turns out that only magnetic gluons of color 1 have a
dispersion branch in this region, cf.\ Figs.\ \ref{fig5} (b) and
\ref{fig6} (b). 
The origin of this branch, the properties of these gluons, and
possible implications for deconfinement in the unbroken $SU(2)_c$ sector 
have been extensively discussed in \cite{rss}. 
Here it is sufficient to note that
for $p_0,\, p \ll \phi$, where the dispersion relation
is approximately linear, the magnetic self-energy
can be written as \cite{rss}
\begin{equation}
\Pi^t_{11}(P)  \simeq  - \frac{m_g^2}{6\, \phi^2}\, p_0^2 \,\, ,
\end{equation}
such that the magnetic
gluon propagator becomes
\begin{equation}
\Delta^t_{11}(p_0,{\bf p}) \simeq \frac{1}{\epsilon\, p_0^2 -  p^2}\,\, ,
\end{equation}
with the dielectricity constant
\begin{equation}
\epsilon = 1 + \frac{m_g^2}{6 \, \phi^2} \,\,.
\end{equation}
The gluon dispersion relation is 
\begin{equation} \label{dispersion}
\omega_{11}^t({\bf p}) \simeq  v\, p \,\, ,
\end{equation}
with the gluon velocity
\begin{equation}
v \equiv \frac{1}{\sqrt{\epsilon}} \simeq \sqrt{6}\, \frac{\phi}{m_g}
\ll 1 \,\, ,
\end{equation}
where the approximate equality and the inequality hold in weak coupling.
The residue along the dispersion relation is
\begin{equation} \label{residue}
Z_{11}^t({\bf p}) \simeq \frac{v}{2\, p} \simeq \sqrt{\frac{3}{2}}\,
\frac{\phi}{m_g\, p}\,\,.
\end{equation}

For the arguments presented in the next section, it is necessary
to consider the limit $\phi \rightarrow 0$ of the
spectral densities. 
In this limit, regions Ia, Ib, and IIb shrink to zero.
In region IIa (which becomes the complete region below the light cone
$p_0 = p$), the regularity of the self-energies
$\Pi^{00}_{11}$ and $\Pi^t_{88}$ implies that the
spectral densities $\rho^{00}_{11}$ and
$\rho^t_{88}$ are also regular. Hence, they converge smoothly
towards the corresponding HDL expressions,
\begin{mathletters} \label{limspecdens}
\begin{eqnarray}
\lim_{\phi \rightarrow 0} \; \rho^{00}_{11}(P) & = & 
\rho^{00}_0(P)\,\, \\
\lim_{\phi \rightarrow 0} \; \rho^t_{88}(P) & = &
\rho^t_0(P)\,\, ,
\end{eqnarray}
for $P $ in region IIa.

As discussed in the previous section,
the real parts of the self-energies $\Pi^t_{11}$ and $\Pi^{00}_{88}$
also converge towards the corresponding HDL expressions
in the limit $\phi \rightarrow 0$, except in a region of size $\sim \phi$
near the logarithmic singularity at $p_0 = 2\, \phi$.
This singularity is rendered harmless in the spectral densities;
the latter simply vanish at this point. The deviation
from the HDL spectral densities is, however, still
large near this point. The limit
$\phi \rightarrow 0$ nevertheless exists, since
the point $p_0 = 2\, \phi$ merges with the point $p_0 = 0$, where
the HDL spectral densities vanish, and the size of the
region where there are deviations simply shrinks to zero. Hence also
\begin{eqnarray}
\lim_{\phi \rightarrow 0} \; \rho^t_{11}(P) & = &
\rho^t_0(P)\,\, ,\\
\lim_{\phi \rightarrow 0} \; \rho^{00}_{88}(P) & = & 
\rho^{00}_0(P)\,\, ,
\end{eqnarray}
\end{mathletters}
for $P $ in region IIa.
Finally, in region III the real parts of {\em all\/}
self-energies converge smoothly towards the HDL expressions when
$\phi \rightarrow 0$. Hence, the dispersion curves in a
color superconductor become identical with the HDL dispersion curves.
Furthermore, the width of the ``smeared'' delta-functions
in Fig.\ \ref{fig6}, being
$\sim {\rm Im}\, \Pi \sim \phi^2$, 
goes to zero and simultaneously their height becomes infinite: 
they become true delta-functions,
because
\begin{equation}
\lim_{\delta \rightarrow 0} \; \frac{\delta/\pi}{x^2 + \delta^2}
\end{equation}
is a representation of $\delta(x)$.
[To prove this, use Eq.\ (\ref{formula}).]
Thus, Eqs.\ (\ref{limspecdens}) also hold in region III, and
hence in all regions of the $(p_0,p)$-plane which have
nonzero measure in the limit $\phi \rightarrow 0$.

\section{The effect on the solution of the gap equation} \label{iv}

We are now in a position to assess the effect of the modification
of the gluon self-energies in a color superconductor on the
solution of the gap equation. Let us first perform
the Dirac algebra in Eq.\ (\ref{gapequation}); details 
can be found in \cite{rdpdhr2}.
After projection onto positive and negative
energy, as well as left- and right-handed particle states, and neglecting
the contribution from the antiparticle gap,
the result for the particle gap function is
\begin{eqnarray}
\phi^+(K) & \simeq & \frac{3}{4}\, g^2 \frac{T}{V} \sum_Q
\frac{\phi^+(Q)}{q_0^2 - [\epsilon_q^+(\phi^+)]^2}\,
\left\{ \left[ \, \Delta^{00}_{11} (K-Q) - \frac{1}{9}\,
\Delta^{00}_{88} (K-Q) \frac{}{} \right]\, \frac{1+\hat{\bf k} \cdot
\hat{\bf q}}{2} \right. \nonumber \\
&     & + \left.  \left[ \, \Delta^t_{11} (K-Q) - \frac{1}{9}\,
\Delta^t_{88} (K-Q) \frac{}{} \right]\, \left[
- \frac{3-\hat{\bf k} \cdot \hat{\bf q}}{2} +  \frac{1+\hat{\bf k} \cdot
\hat{\bf q}}{2} \, \frac{(k-q)^2}{({\bf k} - {\bf q})^2} \right] \right\}
\,\, .
\end{eqnarray}
Here, $\epsilon_q^+(\phi^+) \equiv \sqrt{(q-\mu)^2 + (\phi^+)^2}$,
and the chirality index $r,\, \ell$ of the gap function was omitted,
since left- and right-handed gaps decouple and obey the same gap equation.
To simplify the notation, in the following I shall also omit the 
energy index of the gap function, as well as the energy index and
the argument of the excitation energy $\epsilon_q^+(\phi^+)$.

Now approximate the terms involving $\hat{\bf k} \cdot \hat{\bf q}$ as
\begin{equation}
\frac{1+\hat{\bf k} \cdot \hat{\bf q}}{2} \simeq 1\;\;\;\; ,\;\;\;\;\;
- \frac{3-\hat{\bf k} \cdot \hat{\bf q}}{2} +  \frac{1+\hat{\bf k} \cdot
\hat{\bf q}}{2} \, \frac{(k-q)^2}{({\bf k} - {\bf q})^2} \simeq - 1\,\, ,
\end{equation}
which is appropriate to leading and subleading order in weak coupling.
In the resulting equation,
add and subtract terms with HDL gluon propagators,
\begin{eqnarray} \label{gapequation2}
\phi(K) & \simeq & \frac{2}{3} \, 
g^2 \frac{T}{V} \sum_Q \frac{\phi(Q)}{q_0^2 - \epsilon_q^2}\,
\left[ \; \Delta^{00}_0 (K-Q) - \Delta^t_0(K-Q) \frac{}{} \right]
+ \delta \phi(K)\,\, , \\ 
\delta \phi(K) & \equiv & \frac{3}{4} \, 
g^2 \frac{T}{V} \sum_Q \frac{\phi(Q)}{q_0^2 - \epsilon_q^2}\,
\left\{  \Delta^{00}_{11} (K-Q) - \Delta^{00}_0(K-Q)
- \frac{1}{9}\, \left[\; \Delta^{00}_{88}(K-Q) - \Delta^{00}_0(K-Q)
\frac{}{} \right] \right. \nonumber \\
&    & \hspace*{2.55cm}
- \left. \left[\; \Delta^t_{11} (K-Q) - \Delta^t_0(K-Q) \frac{}{} \right]
+ \frac{1}{9}\, \left[\; \Delta^t_{88}(K-Q) - \Delta^t_0(K-Q)
\frac{}{} \right] \right\}
\,\, .
\end{eqnarray}
For $\delta \phi(K) =0$, we
recover the standard gap equation with the gluon
propagators in HDL approximation. To leading
and subleading order in weak coupling, this equation is
solved by the gap function derived by Son \cite{son}. At the Fermi
surface, the value of the gap is given by Eq.\ (\ref{barroisresult}) 
with $c$ and $b$ from Eqs.\ (\ref{sonresult1}) and (\ref{b}), respectively.
The additional term $\delta \phi(K)$ 
is the correction to this solution. In the following,
drawing on the results of the previous sections I determine 
whether this correction enters at leading order
(modifying the value of $c$), subleading order (modifying the
value of $b$), or sub-subleading order (no modification of $c$ and $b$
at all).

First, rewrite the gluon propagators in Eq.\ ({\ref{gapequation})
in terms of spectral densities, see also Eq.\ (33) of \cite{rdpdhr2}, 
\begin{mathletters}
\begin{eqnarray}
\Delta^{00}(P) & \equiv & - \frac{1}{p^2} + \int_0^{1/T} {\rm d}\tau\,
e^{p_0 \tau}\, \int_0^\infty {\rm d} \omega\, \rho^{00}(\omega,{\bf p})\;
\left\{ \frac{}{} \left[1+n_B(\omega/T) \right]\,e^{-\omega \tau}
+ n_B(\omega/T) \, e^{\omega \tau} \; \right\} \,\, , \\
\Delta^t(P) & \equiv & \int_0^{1/T} {\rm d}\tau\,
e^{p_0 \tau}\, \int_0^\infty {\rm d} \omega\, \rho^t(\omega,{\bf p})\;
\left\{ \frac{}{} \left[1+n_B(\omega/T) \right]\,e^{-\omega \tau}
+ n_B(\omega/T) \, e^{\omega \tau} \; \right\} \,\, .
\end{eqnarray}
\end{mathletters}
Here, I momentarily reverted to nonzero temperature until the
Matsubara sum over fermionic energy $q_0 = -i (2n+1) \pi T$ in
Eq.\ (\ref{gapequation2}) has been performed. Consequently,
$n_B(x) \equiv \left( e^x -1 \right)^{-1}$ is the Bose-Einstein
distribution function. The term $-1/p^2$ in the electric
propagator cancels the contribution of $\Delta^{00}(P)$ when
$p_0 \rightarrow \infty$ \cite{LeBellac}. This term
is the same for the electric gluon propagators in a superconductor 
and the electric HDL propagator, because, as we have seen 
in the previous sections, there is no difference
between these propagators for $p_0 \gg \phi$.

To proceed, one also needs the spectral representation 
\begin{equation}
\frac{\phi(Q)}{q_0^2 - \epsilon_q^2} = -
\int_0^{1/T} {\rm d} \tau \, e^{q_0 \tau}\,
\frac{\phi(\epsilon_q,{\bf q})}{2 \epsilon_q}
\left\{\; \left[ 1 - n_F(\epsilon_q/T) \right]\,
e^{- \epsilon_q \tau}
- n_F(\epsilon_q/T) \, e^{\epsilon_q \tau} \frac{}{} \right\}\,\, ,
\end{equation}
where $n_F(x) \equiv (e^x +1)^{-1}$ is the Fermi-Dirac distribution
function. The Matsubara sum over $q_0$ can now be performed in
the standard manner. After analytic continuation onto the
quasiparticle mass shell, $k_0 \rightarrow \epsilon_k + i \eta$,
and abbreviating $\phi(\epsilon_q,{\bf q}) \equiv \phi_q$
the equation for $\delta \phi_k$ reads
\begin{eqnarray}
\delta \phi_k & = &- \frac{3}{4}\, g^2 \int 
\frac{{\rm d}^3 {\bf q}}{(2 \pi)^3}\, \frac{\phi_q}{2 \epsilon_q}
\int_0^\infty {\rm d} \omega
\left[ \delta \rho^{00}_{11}(\omega,{\bf p}) - \frac{1}{9} \,
\delta \rho^{00}_{88}(\omega,{\bf p}) - \delta \rho^t_{11}
(\omega,{\bf p}) + \frac{1}{9}\, \delta \rho^t_{88}(\omega,{\bf p})
\right] \nonumber \\ 
&     & \hspace*{3.9cm}
\times \left( \frac{1}{\omega + \epsilon_q + \epsilon_k} +
\frac{1}{\omega + \epsilon_q - \epsilon_k} \right)\,\, , \label{64}
\end{eqnarray}
where ${\bf p} \equiv {\bf k} - {\bf q}$, and where I introduced
\begin{equation}
\delta \rho^{00,\,t}_{aa} \equiv \rho^{00,\,t}_{aa} - \rho^{00,\, t}_0
\;\;\;\; , \;\;\;\; a = 1,\, 8\,\, .
\end{equation}
Neglecting a possible imaginary part of the gap function
(cf.\ discussion in \cite{rdpdhr2}), it is implicitly assumed
that the energy denominators in Eq.\ (\ref{64})
are evaluated with the principal value prescription.

Finally, note that the spectral densities are isotropic, {\it i.e.},
they only depend on $|{\bf p}| \equiv p$. Substituting the
integration variable $\cos \theta \equiv \hat{\bf k} \cdot \hat{\bf q}$
by $p$, and interchanging the order in which the $p$- and $\omega$-integrals
are performed, one obtains
\begin{equation} \label{deltaphi}
\delta \phi_k \simeq \frac{3\, g^2}{32 \pi^2}  \int_{\mu-\delta}^{\mu+\delta}
{\rm d} q \, \frac{\phi_q}{\epsilon_q} \int_0^\infty {\rm d}
\omega \left\{ {\cal J}^t_{11}(\omega) -{\cal J}^{00}_{11} (\omega)
 - \frac{1}{9} \left[ {\cal J}^t_{88}(\omega) - {\cal J}^{00}_{88}
(\omega) \right] \right\} \left( \frac{1}{\omega + \epsilon_q + \epsilon_k} +
\frac{1}{\omega + \epsilon_q - \epsilon_k} \right) \, .
\end{equation}
where I  defined
\begin{equation} \label{J}
{\cal J}^{00,\,t}_{aa} (\omega) \equiv \int_0^{2\mu} {\rm d} p \, p \;
\delta \rho^{00,\,t}_{aa} (\omega,p)\,\,.
\end{equation}
In deriving Eq.\ (\ref{deltaphi}) with Eq.\ (\ref{J}) I
also approximated $k/q \simeq 1$, $k+q \simeq 2 \mu$, $|k-q| \simeq 0$,
which is valid, since the $q$-integral peaks at the Fermi surface, 
see also \cite{rdpdhr2}. Moreover, since the gap function $\phi_q$
depends strongly on $q$ \cite{son}, it is permissible to restrict
the momentum integration to a small region around
the Fermi surface, $\mu-\delta \leq q \leq \mu + \delta$
where $\delta \ll \mu$.

The important point is that the functions ${\cal J}^{00,\,t}_{aa}$
are {\em dimensionless}, {\it i.e.}, they can only be functions of
the dimensionless variables $\omega/m_g$ and $\phi/m_g$ (or $\omega/\mu$
and $\phi/\mu$; additional factors of $g$ turn out to be of no consequence
for the following argument).
Let us now estimate these functions.

\subsection{Estimate for ${\cal J}^{00}_{11}$}

For $\omega \leq 2\, \phi$, $\rho^{00}_{11} \equiv 0$, and
$\delta \rho^{00}_{11} = - \rho^{00}_0$.
When computing the $p$-integral in ${\cal J}^{00}_{11}$, 
one samples $\rho^{00}_0$ in regions Ia and
Ib of Fig.\ \ref{fig4}. From the explicit form of the spectral density
\cite{rdpdhr2,LeBellac,RDPphysica},
\begin{equation}
\rho^{00}_0(\omega,p) = \theta(p-\omega)\,
\frac{2M^2}{\pi}\, \frac{\omega}{p}\,
\left\{ \left[ p^2+3\, m_g^2 \, \left( 1 -\frac{\omega}{2p} \, 
\ln \left| \frac{p+\omega}{p-\omega} \right| \right) \right]^2 + 
\left(2 M^2 \,\frac{\omega}{p} \right)^2 \right\}^{-1}\,\, ,
\end{equation}
where $M^2 \equiv 3 \pi \, m_g^2/4$, one deduces that
$\rho^{00}_0$ vanishes in region Ib, while in region Ia $\rho^{00}_0$ peaks
for $p$ of order (and slightly larger than) $\omega$, and then falls 
off $\sim 1/p^5$. Neglecting powers of $g$, 
${\cal J}^{00}_{11}$ is at most
\begin{equation} \label{63}
{\cal J}^{00}_{11} (\omega) \sim \frac{\omega}{m_g} \sim \frac{\phi}{m_g}\,\, .
\end{equation}

For $\omega > 2\, \phi$, the $p$-integral samples regions III, IIb, and
IIa. In the latter, both $\rho^{00}_{11}$ and $\rho^{00}_0$
are regular, consequently $\delta \rho^{00}_{11}$ is regular, and
from the discussion in the previous section one concludes that it 
vanishes in the limit $\phi \rightarrow 0$. In principle, one
could therefore Taylor-expand $\delta \rho^{00}_{11}$ around
$\phi = 0$. To leading order, one then has in region IIa
\begin{mathletters} \label{cases}
\begin{equation}
\int_\omega^{2\mu} {\rm d} p \, p \; 
\delta \rho^{00}_{11}(\omega,p) \sim \frac{\phi}{m_g}\,\, .
\end{equation}
In region IIb, $\rho^{00}_0 = 0$, while $\rho^{00}_{11}$ is regular
and of order $1/m_g^2$. The $p$-integral over this region is therefore
at most
\begin{equation}
\int_{\sqrt{\omega^2-4\phi^2}}^\omega {\rm d} p\, p \,
\delta \rho^{11}_{00}(\omega,p) \sim \frac{\phi^2}{m_g^2}\,\,.
\end{equation}
Finally, in region III, $\delta \rho^{11}_{00}$ is the difference
between a ``smeared'' delta-function and a true delta-function, which
is singular.
However, the $p$-integral over $\delta \rho^{11}_{00}$ 
renders this singularity regular, and since $\delta \rho^{11}_{00}$ 
vanishes in the limit
$\phi \rightarrow 0$ also in this region, the same
argument as in region IIa applies,
\begin{equation}
\int_0^{\sqrt{\omega^2 - 4 \phi^2}} {\rm d} p \, p \;
\delta \rho^{11}_{00}(\omega,p) \sim \frac{\phi}{m_g}\,\, .
\end{equation}
\end{mathletters}
Summarizing Eqs.\ (\ref{63}) and (\ref{cases}), one concludes
\begin{equation}
{\cal J}^{00}_{11} (\omega) \sim \frac{\phi}{m_g}
\end{equation}
for all $\omega$.

\subsection{Estimate for ${\cal J}^t_{11}$}

The estimate for the function ${\cal J}^t_{11}$ is similar.
The difference is that, in region Ia, there is also a contribution
from $\rho^t_{11}$. Explicitly,
\begin{equation}
\int_\omega^{2\mu} {\rm d}p \, p \; \rho^t_{11}(\omega,p)
= - p(\omega_{11}^t) \, Z_{11}^t(p) \, 
\left( \left| \frac{\partial \omega_{11}^t(p)}{\partial p} \right|
\right)^{-1}_{p = p(\omega_{11}^t)} \,\, ,
\end{equation}
where the last term is the Jacobian from the argument
of the delta-function in Eq.\ (\ref{singular}).
With Eqs.\ (\ref{dispersion}) and (\ref{residue}) one concludes
\begin{equation} \label{nontrivial}
\int_\omega^{2\mu} {\rm d}p \, p \; \rho^t_{11}(\omega,p)
\sim 1 \,\, .
\end{equation}
For our order-of-magnitude estimate, it is sufficient
to use the approximate form of the magnetic HDL spectral density
\cite{rdpdhr2,RDPphysica}
\begin{equation} \label{approximaterhot}
\rho^t_0(\omega,p) \simeq \theta(p - \omega)\, 
\frac{M^2}{\pi}\, \frac{\omega\, p}{p^6 + (M^2 \omega)^2}\,\, ,
\end{equation}
valid for $p \gg \omega$ (this is where the magnetic HDL
spectral density has its maximum).
One then estimates
\begin{equation}
\int_0^{2\mu} {\rm d} p \, p \; \rho^t_0(\omega,p) \sim
\frac{\omega}{m_g} \sim \frac{\phi}{m_g} \,\, .
\end{equation}
Thus, to leading order
\begin{equation}
{\cal J}^t_{11}(\omega) = \int_0^{2\mu} {\rm d} p \, p \; 
\delta \rho^t_{11}(\omega,p) \sim 1 \,\, .
\end{equation}

For $\omega > 2\, \phi$, the situation is similar as for
${\cal J}^{00}_{11}$ and I refrain from repeating the details.
One might worry that near $\omega = 2 \, \phi$, where the deviation
of $\rho^t_{11}$ from $\rho^t_0$ is
large due to the logarithmic singularity of 
${\rm Re}\, \Pi^t_{11}$, the above arguments do
not hold. However, here $\delta \rho^t_{11} \simeq - \rho^t_0$
and with Eq.\ (\ref{approximaterhot})
one convinces oneself that still
${\cal J}^t_{11} \sim \phi/m_g$.
In conclusion, to leading order
\begin{equation}
{\cal J}^t_{11}(\omega) \sim \theta (2 \, \phi - \omega)\, \times \, 1
        + \theta(\omega - 2\, \phi) \, \times \, \frac{\phi}{m_g} \,\, .
\end{equation}

\subsection{Estimate for ${\cal J}^{00}_{88}$ and ${\cal J}^t_{88}$}

The functions ${\cal J}^{00}_{88}$ and ${\cal J}^t_{88}$ are
easy to estimate. 
For $\omega \leq 2 \, \phi$, $\delta \rho^{00,\, t}_{88}$
is the difference between a spectral density with $2/3$ of the value of
the HDL self-energies and the HDL spectral density itself.
Following the same line of arguments as in the previous cases,
${\cal J}^{00,\, t}_{88}(\omega) \sim \phi/m_g$.
For $\omega > 2\, \phi$, the situation is similar to that for the
functions ${\cal J}^{00,\, t}_{11}$.
This immediately leads to the estimate
\begin{equation}
{\cal J}^{00, \,t}_{88}(\omega) \sim \frac{\phi}{m_g}\,\, .
\end{equation}

\subsection{Estimate for $\delta \phi_k$}

We are now in the position to estimate $\delta \phi_k$.
With the exception of ${\cal J}^t_{11}$ in the region $\omega \leq
2 \, \phi$, all functions ${\cal J}^{00,\, t}_{aa}$ contribute
at most terms of order $\phi/m_g$ to $\delta \phi_k$.
These contributions are suppressed by 
$\phi/m_g$ relative to the leading or subleading
contributions.
Additional powers of $g$ are of no consequence, because
they are accompanied by at least one power of $\phi$ which
is exponentially small in $g$, $\phi \sim \exp(-1/g)$.
Thus, these contributions cannot change the values of the constants 
$c$ and $b$ of Eqs.\ (\ref{sonresult1}) and (\ref{b}).

The only remaining contribution with a potential impact
on the values for $c$ or $b$ is the one from ${\cal J}^t_{11}$.
This contribution gives rise to a $\delta \phi_k$ which is of order
\begin{equation} \label{FINALESTIMATE}
\delta \phi_k \sim g^2  \int_{\mu-\delta}^{\mu+\delta} {\rm d} q \;
\frac{\phi_q}{\epsilon_q} \; \ln \left|
\frac{(2\, \phi + \epsilon_q)^2 - \epsilon_k^2}{\epsilon_q^2 - \epsilon_k^2} 
\right|\,\, .
\end{equation}
The integral can be estimated to be $\sim \phi$, cf.\ Appendix
\ref{app3},
such that $\delta \phi_k \sim g^2 \phi$. This means that
$\delta \phi_k$ contributes at most to the constant $d$
in Eq.\ (\ref{QCDgapequation}), {\it i.e.}, to
sub-subleading order to the solution of the gap equation.
Thus, to leading and subleading order, the solution
(\ref{barroisresult}) with (\ref{sonresult1}) and (\ref{b})
is not modified.

\section{Summary, conclusions, and outlook} \label{v}

In this paper, I have computed the electric and magnetic
self-energies for gluons of adjoint colors 1, 2, 3, and 8
in a color superconductor with two flavors of massless quarks. 
In an extension of a previous study \cite{dhr2f} which was
restricted to gluon energies and momenta $p_0 = p = 0$, as well
as $p_0 = 0$, $\phi \ll p \ll \mu$, I now considered the complete
range $p_0,\, p \ll \mu$.
The results can be qualitatively summarized as follows: the
presence of a color-superconducting condensate
modifies the gluon self-energies as compared to the self-energies
in a normal-conducting, dense medium only for
energies $p_0 \sim \phi$.
For $p_0 \gg \phi$, the expressions for the 
self-energies approach the standard hard-dense-loop results.

This study was motivated by the following two observations.
First, the weak-coupling result for the color-superconducting gap parameter 
depends sensitively on the number of unscreened ({\it i.e.} massless)
gluon degrees of freedom.
Second, in a two-flavor color superconductor not only magnetic but also
electric gluons are massless in the static, homogeneous limit \cite{dhr2f}.
This led to the speculation as to whether the modification of the
gluon propagator in a color superconductor could change the value
of the gap parameter to subleading order in weak
coupling, Eq.\ (\ref{barroisresult}) with (\ref{sonresult1}) and (\ref{b}).
Using the results for the gluon self-energies, I 
estimated that this is not the case. The physical reason is that the
change of gluon properties as compared to those in a normal
conductor is limited to the rather small region $p_0 \leq \phi$.

There remains one unanswered question:
the peculiar behavior of the dispersion relation of electric
gluons with color 8 displayed in Fig.\ \ref{fig5} (a).
As discussed above, this could be due to the fact that the present calculation
neglects the mixing of this
gluon with the excitations of the condensate. This will be addressed
in greater detail in a forthcoming publication \cite{shovrisch}.
Since this phenomena is restricted to momenta $p \sim \phi$, it is, however, 
unlikely that the
conclusions regarding the stability of the value of the gap parameter
to leading and subleading order will have to be revised.

\section*{Acknowledgements}

I am indebted to D.\ Boyanovsky, G.\ Carter, 
C.\ Manuel, K.\ Rajagopal, I.\ Shovkovy, M.\ Stephanov,
D.\ Son, and especially R.\ Pisarski for stimulating discussions. I thank
J.\ Knoll for discussions on the contents of Appendix \ref{app2}.
G.\ Carter, R.\ Pisarski, T.\ Sch\"afer, I.\ Shovkovy, and D.\ Son are
gratefully acknowledged for critically reviewing the manuscript.
I thank my colleagues at the RIKEN-BNL Research Center and in the
Nuclear and High Energy Theory groups at BNL 
for their hospitality and friendship extended
towards me during the past three and a half years.
I also thank RIKEN, BNL, and the U.S.\ Dept.\ of Energy for
providing the facilities essential for the completion of this work,
and Columbia University's Nuclear Theory Group for
continuing access to their computing facilities.

\appendix

\section{Numerical calculation of the real part of the gluon self-energy}
\label{app2}

The real parts of the gluon self-energies can either be computed
from Eqs.\ (\ref{Pi11all}), (\ref{Pi88all}) with Eq.\ (\ref{formula}),
or from the dispersion integral (\ref{dispersionintegral}).
In both cases, this amounts to evaluating a double integral.
In the first case, this integral runs over $\xi=k-\mu$ and $x= \cos \theta$, 
while in the second case, one has to compute
elliptic integrals numerically in addition to the integral over $\omega$.

The second way is, however, the simpler one.
As the integrand falls off $\sim 1/\omega^3$, one only needs
to compute the integral up to some finite value $\Omega$
which is sufficiently large for the required numerical accuracy.
One then divides the interval $[0,\Omega]$ into
$N$ pieces of size $\delta \Omega = \Omega/N$, 
such that $\omega_n = n \, \delta \Omega$, $n=0,\ldots, N$, and
computes the integral over $\omega$ piecewise 
with a generalization of the mean-value theorem as
\begin{equation}
{\rm Re}\, \Pi(p_0,{\bf p}) \simeq \frac{1}{\pi}\; \sum_{n=0}^N 
{\rm Im}\, \Pi(\omega^*_n,{\bf p})\; \ln \left| 
\frac{\omega_{n+1}^2 - p_0^2}{\omega_n^2 - p_0^2} \right| + C \,\, ,
\end{equation}
with $\omega_n^* \in [\omega_n,\omega_{n+1}]$ (for practical
purposes, $\omega_n^* \equiv (\omega_{n+1} + \omega_n)/2$ is
a convenient choice). In agreement with the principal-value prescription
in (\ref{dispersionintegral}), one has to make sure to avoid
$\omega_n = p_0$ for any $n$.
The advantage of this method is that
the imaginary parts can be tabulated prior to the computation of
the real parts, such that the computation of the double integral is
effectively reduced to the computation of two independent
ordinary integrals.

\section{Estimate of the integral in Eq.\ (\ref{finalestimate})} \label{app3}

In this Appendix, I estimate the value of the integral in
Eq.\ (\ref{FINALESTIMATE}),
\begin{equation}
{\cal I}(\epsilon_k) \equiv 
\int^{\mu+\delta}_{\mu - \delta} {\rm d}q \, \frac{\phi_q}{\epsilon_q} 
 \; \ln \left| 
\frac{(2\, \phi + \epsilon_q)^2 - \epsilon_k^2}{\epsilon_q^2 - \epsilon_k^2} 
\right|\,\,.
\end{equation}
In order to see whether $\delta \phi_k$ contributes to
the constants $b,\,c,$ or $d$ in Eq.\ (\ref{QCDgapequation}), it is sufficient
to perform this estimate at the Fermi surface, $k = \mu$, {\it i.e.},
$\epsilon_k = \phi$.
Since the gap function is peaked at
the Fermi surface, $\phi_q \leq \phi_\mu \equiv \phi$,
\begin{equation}
{\cal I}(\phi) \leq \phi 
\int^{\mu+\delta}_{\mu - \delta} \frac{{\rm d}q}{\epsilon_q} 
 \; \ln \left[ 
\frac{(2\, \phi + \epsilon_q)^2 - \phi^2}{\epsilon_q^2 - \phi^2} 
\right] \equiv 2\, \phi 
\int_0^\delta \frac{{\rm d} \xi }{\epsilon_q}
\; \ln \left[ 
\frac{(2\, \phi + \epsilon_q)^2 - \phi^2}{\epsilon_q^2 - \phi^2} 
\right]\,\, ,
\end{equation}
where $\xi \equiv q - \mu$.
Now substitute the integration variable
\cite{rdpdhr2}
\begin{equation}
y \equiv \ln \frac{\xi+ \epsilon_q}{\phi}\,\, ,
\end{equation}
with the result
\begin{equation}
{\cal I}(\phi) \leq 2\, \phi \int^{\ln (2 \delta/\phi)}_0
{\rm d}y \; \ln \left[ \frac{\cosh^2 (y/2) + 1}{\sinh^2 (y/2)} \right] \,\, .
\end{equation}
The right-hand side of this equation can be further evaluated,
\begin{equation}
{\cal I}(\phi) \leq 2\, \phi\, \left\{
2 \int^1_{\phi/(2 \delta)} \frac{{\rm d}x}{x}\; \ln \frac{1+x}{1-x}
+ \int^{\ln (2 \delta/\phi)}_0 {\rm d} y \; \ln \left[
1 + \frac{1}{\cosh^2 (y/2)} \right] \right\} \,\, ,
\end{equation}
where $x \equiv e^{-y}$. An upper bound for the first integral
is given by extending the lower boundary to zero. The resulting 
integral can be solved analytically and has the value $\pi^2/4$.
An upper bound for the second integral is given by extending
the upper boundary to infinity.
An integration by parts transforms the resulting integral into
\begin{equation}
{\cal I}'
\equiv \int_0^\infty {\rm d}y \; y \; \frac{\tanh(y/2)}{\cosh^2(y/2) + 1}
\,\, .
\end{equation}
Substituting $x \equiv \cosh(y/2)$,
\begin{equation} \label{Iprime}
{\cal I}' = 4 \int_1^\infty \frac{{\rm d}x}{x(1 + x^2)}\;
\ln \left( x + \sqrt{x^2-1}\right) 
\leq 4 \int_1^\infty \frac{{\rm d}x}{x(1 + x^2)}\; \ln ( 2 x)\,\, ,
\end{equation}
as $\sqrt{x^2-1} \leq x$ for $x \in [1,\infty)$.
The last integral in (\ref{Iprime}) can be solved analytically.
Collecting the results, the integral ${\cal I}(\phi)$ is bounded
from above by
\begin{equation}
{\cal I}(\phi) \leq 2\, \phi \left[ \frac{7 \pi^2}{12} + 2\, (\ln 2)^2
\right]\,\, .
\end{equation}
In conclusion, ${\cal I}(\phi)$ is of order $const. \times \phi$,
and thus $\delta \phi_k \sim g^2  \phi$.

\end{document}